\documentclass[11pt,twocolumn]{article}
\usepackage{etoolbox}     
\usepackage{orcidlink}
\usepackage{amsmath}
\usepackage{amsfonts}
\usepackage{amssymb,amsthm}
\usepackage{graphicx}
\usepackage{orcidlink}
\usepackage{hyperref}
\hypersetup{colorlinks=true, linkcolor=blue, citecolor=blue, urlcolor=blue}
\usepackage{longtable}
\usepackage{booktabs}
\usepackage{colortbl}
\usepackage{array}
\usepackage{placeins}
\RequirePackage[numbers,sort&compress]{natbib}
\begin{document}
\baselineskip=14pt

\twocolumn[
\begin{center}
\large{\bf Quantum-Corrected Thermodynamics of Conformal Weyl Gravity Black Holes: \\[1ex] GUP Effects and Phase Transitions}
\end{center}
\vspace{0.3cm}
\begin{center}
{\bf Erdem Sucu\orcidlink{0009-0000-3619-1492}}\footnote{\bf erdemsc07@gmail.com}\\
{\it Eastern Mediterranean University, Physics Department, Famagusta, 99628 North Cyprus, via Mersin 10, T\"urkiye}\\
{\bf Suat Dengiz\orcidlink{0000-0002-7099-4608}}\footnote{\bf suat.dengiz@atilim.edu.tr}\\
{\it Department of Electrical and Electronics Engineering, Faculty of Engineering, At{\i}l{\i}m University, 06836 Ankara, T\"urkiye}\\
{\bf \.{I}zzet Sakall{\i}\orcidlink{0000-0001-7827-9476}}\footnote{\bf izzet.sakalli@emu.edu.tr (Corresp. author)}\\
{\it Eastern Mediterranean University, Physics Department, Famagusta, 99628 North Cyprus, via Mersin 10, T\"urkiye}
\end{center}
\vspace{0.3cm}
\begin{center}
\begin{minipage}{0.92\textwidth}
\noindent\textbf{Abstract.} We investigate the thermodynamic properties of black holes in Conformal Weyl Gravity (CWG) using the Mannheim-Kazanas solution, with particular emphasis on quantum corrections that become significant near the Planck scale. Our analysis employs the Hamilton-Jacobi tunneling formalism to derive the Hawking temperature, revealing explicit contributions from the conformal parameters $\beta$, $\gamma$, and $k$ that lead to substantial deviations from Schwarzschild black hole behavior. We incorporate quantum gravitational effects through the Generalized Uncertainty Principle, demonstrating systematic suppression of thermal radiation in the near-Planckian regime. Using an exponentially corrected entropy model, we compute the complete spectrum of QC thermodynamic potentials, including internal energy, pressure, heat capacity, and free energies. Our heat capacity analysis shows divergence behavior that separates stable and unstable regions, indicating possible thermodynamic transitions controlled by the scale-dependent parameter $\gamma$. The Joule-Thomson expansion analysis shows distinct cooling and heating regimes with inversion points that shift systematically with CWG parameters, capturing QC phase transitions absent in general relativity. We also examine gravitational redshift in CWG geometry, finding complex radial dependence that highlights modifications compared to the Schwarzschild case, although redshift alone cannot observationally distinguish CWG from Einstein's theory. Our results demonstrate that CWG offers a consistent framework for studying black hole thermodynamics beyond general relativity, with quantum corrections modifying phase structures in the near-Planckian regime, though these effects are not expected to yield direct observational consequences.

\vspace{0.3cm}
\noindent\textbf{Keywords:} Black hole; Joule-Thomson Expansion; Quantum Correction; Redshift; Radiation; Generalized Uncertainty Principle; Conformal Weyl Gravity.
\end{minipage}
\end{center}
\vspace{0.6cm}
]

\section{Introduction}\label{isec1}

The study of black hole (BH) thermodynamics has emerged as one of the most profound and intellectually rewarding intersections between general relativity (GR), quantum field theory, and statistical physics, fundamentally transforming our understanding of the deep connections between gravity, thermodynamics, and quantum mechanics. Although Einstein's GR has achieved remarkable success in describing macroscopic gravitational phenomena across vast effective energy scales, from planetary motion to cosmological dynamics, several fundamental conceptual and observational challenges at both infrared and ultraviolet scales have persistently motivated the exploration of alternative theoretical frameworks. These challenges include the notorious cosmological constant problem, the enigmatic galactic rotation curve anomalies that suggest the presence of dark matter, the quantum origin of the BH entropy and the associated information paradox, and the fundamental failure of GR to provide a complete quantum theory of gravity \cite{shankaranarayanan2022modified,stoica2018revisiting}.

CWG stands out prominently among alternative gravity theories due to its elegant local conformal invariance and sophisticated higher-derivative mathematical structure \cite{Mannheim:1988dj,Mannheim:2011ds, Mannheim:1999bu, Mannheim:1989jh, Mannheim:2006rd}. Unlike conventional GR, which constructs its gravitational action from the Ricci scalar $R$, CWG employs the square of the conformal Weyl tensor $C_{\mu\nu\rho\sigma}C^{\mu\nu\rho\sigma}$ as its fundamental building block, thereby introducing fourth-order field equations that admit significantly richer vacuum structures and more complex dynamical behaviors. The Mannheim-Kazanas (MK) solution, derived within this theoretical framework, represents a notable departure from classical Schwarzschild geometry through the inclusion of additional linear $\gamma r$ and quadratic $k r^2$ potential terms that emerge naturally from higher-order field equations \cite{Mannheim:1988dj,Mannheim:2005bfa}. These non-Einsteinian contributions offer compelling explanations for various astrophysical observations, including galactic rotation curves and large-scale structure formation, without requiring the introduction of exotic dark matter components \cite{mannheim2012fitting}. Furthermore, the presence of local conformal symmetry in CWE does not tolerate any dimensionful parameter (and hence no rigid scale) at sufficiently high energy scales. So, the Newton constant with dimension $[m^{-2}]$-which causes the irremediable catastrophic infinities at loop-level and hence leads to perturbatively \emph{non}-renormalizable quantum gravity-comes into existence only in the classical vacuum of CWE where the conformal symmetry is broken.

The thermodynamic properties of BHs within the CWG exhibit fascinating modifications compared to their GR counterparts \cite{Ahmed:2025sav,Ahmed:2025qow,Al-Badawi:2025rya,Ahmed:2025wmt,Al-Badawi:2025rcq}, leading to altered causal structures, modified thermal radiation spectra, and fundamentally different horizon properties that reflect the underlying conformal symmetry \cite{sakti2024thermodynamics,xu2017black}. These modifications become particularly pronounced in extreme parameter regimes, where the additional CWG terms dominate over the standard Newtonian contributions, suggesting potential observational signatures that could distinguish CWG from conventional gravity theories through precision astronomical measurements.

The incorporation of quantum corrections into BH thermodynamics represents another crucial frontier in our understanding of gravitational physics near the Planck scale, where classical descriptions inevitably break down and quantum effects become dominant \cite{calmet2021quantum,campos2022quantum}. The Generalized Uncertainty Principle (GUP) \cite{Gecim:2019pft,WOS:001617915000001,pourhassan2020pv,sucu2025charged,tekincay2021zitterbewegung,kanzi2019gup,tekincay2021exotic,sakalli2025zitterbewegung,Adler:2001vs,sucu2025scalar}, which emerges naturally from various approaches to quantum gravity including string theory, loop quantum gravity, and non-commutative geometry, provides a minimal yet powerful framework for capturing leading-order quantum gravitational effects through the introduction of a fundamental minimal length scale \cite{pedram2011effects,gecim2017gupp,nozari2010minimal,gecim2018quantum,nozari2012minimal,gecim2020quantum,sucu2025quantumRoyal}. This principle modifies the standard Heisenberg uncertainty relations and leads to systematic corrections in BH thermal properties, particularly affecting temperature, entropy, and heat capacity in ways that become increasingly important as BH masses approach Planck-scale values.

Recent advances in observational astronomy, particularly through gravitational wave detection by LIGO/Virgo collaborations and direct BH imaging by the Event Horizon Telescope, have opened unprecedented opportunities for testing alternative gravity theories and probing quantum gravitational effects in realistic astrophysical contexts \cite{vagnozzi2023horizon,akiyama2022first}. These observational breakthroughs demand corresponding theoretical developments that can provide precise predictions for observable signatures of modified gravity and quantum corrections in BH physics.

The Joule-Thomson expansion (JTE) represents a particularly sensitive probe of BH thermodynamics, providing detailed information about phase transitions and thermal stability properties that are invisible to other thermodynamic measures \cite{mo2018joule,okcu2017joule,cisterna2019joule,okcu2018joule,aydiner2025regular}. In the extended phase space formalism, where the cosmological constant is treated as a thermodynamic pressure, JTE analysis reveals rich phase structures including cooling and heating regimes separated by critical inversion points. These phenomena become even more complex in CWG due to the additional parameter space introduced by the conformal corrections. Gravitational redshift measurements offer another powerful avenue for testing alternative gravity theories, as they provide direct observational access to the spacetime metric structure near massive compact objects \cite{guo2023joule,xu2018thermodynamic}. The modification of redshift patterns in CWG compared to GR predictions could potentially be detected through high-precision spectroscopic observations of radiation emanating from the vicinity of supermassive BHs or neutron stars.

Our primary motivation for this comprehensive investigation stems from the recognition that understanding BH physics in alternative gravity theories, particularly when augmented with quantum corrections, represents a crucial step toward developing a complete theory of quantum gravity. CWG provides an ideal theoretical laboratory for this investigation due to its mathematical elegance, observational relevance, and rich phenomenological structure. Using the MK metric as a gravitational background, we systematically investigate the thermodynamic behavior of BHs within CWG, with particular focus on quantum corrections that become significant near the Planck scale. We first revisit the semiclassical Hawking radiation through the Hamilton-Jacobi tunneling method \cite{parikh2000hawking,sucu2025quantumm,sucu2023gup} and then incorporate corrections from the GUP, which is widely regarded as a minimal extension capturing leading-order quantum gravity effects \cite{adler2001generalized,Nozari:2008rc}. These corrections impact key thermodynamic quantities such as temperature, entropy, and heat capacity, opening possibilities for rich phase structures, including modifications to the JTE and redshift phenomena.

The relevance of this analysis lies in its capacity to offer deeper understanding of BH thermodynamics in a conformally invariant setting, enriched with quantum statistical considerations \cite{ghosh2006counting,krasnov1998quantum,chatterjee2020exponential}. Moreover, by exploring the interplay between higher-derivative gravity and minimal-length scenarios, our study contributes toward the broader goal of probing the semi-quantum regime where conventional gravitational and thermodynamic intuitions begin to break down \cite{cai2010black,hossenfelder2013minimal}. Such insights could prove instrumental in guiding phenomenological models in quantum gravity and in interpreting upcoming observational data from gravitational-wave astronomy and BH imaging \cite{sucu2025quantum,barack2019black,jetzer2022quantum}. The specific objectives of our research include the development of a comprehensive thermodynamic framework for CWGBHs that incorporates both the geometric modifications arising from higher-order curvature terms and the quantum corrections emerging from the GUP. We systematically analyze thermal radiation properties, compute QC thermodynamic potentials including energy, entropy, pressure, and heat capacity, and investigate phase transition phenomena through detailed thermodynamic analysis. Additionally, we explore gravitational redshift modifications that could provide observational signatures of CWG effects in astrophysical contexts.

The paper is organized as follows: In Section~\ref{isec2}, we provide a comprehensive review of the MK BH geometry within the CWG framework, establishing the mathematical foundation for our subsequent analysis. Section~\ref{isec3} presents our evaluation of Hawking radiation using the Hamilton-Jacobi formalism for tunneling particles, revealing the thermal spectrum modifications introduced by conformal corrections. Section~\ref{isec4} is devoted to the systematic calculation of quantum corrections arising from the GUP and their impact on thermal radiation properties. In Section~\ref{isec5}, we conduct a detailed study of quantum-corrected (QC) thermodynamics for CWGBHs, including the computation of all relevant thermodynamic potentials and their phase structure analysis. Section~\ref{isec6} focuses on the QC JTE in CWGBHs, investigating isenthalpic processes and thermal stability properties. Section~\ref{isec7} analyzes gravitational redshift phenomena in CWGBH geometry, exploring potential observational signatures of conformal corrections. Finally, Section~\ref{isec8} concludes our investigation by summarizing the main results and outlining prospective research directions that emerge from our findings.

\section{Review Of CWGBH Spacetime: Analysis of MK Solution}\label{isec2}

In CWG, BH solutions emerge from a fundamentally different gravitational action compared to Einstein's GR. Rather than being based on the Ricci scalar $R$ as in the conventional Einstein-Hilbert formulation, CWG employs the square of the Weyl tensor $C_{\mu\nu\rho\sigma}$ as its foundational building block. This approach leads to a locally conformally invariant action that takes the elegant form \cite{Mannheim:1988dj, Mannheim:2011ds,Mannheim:1999bu,Mannheim:1989jh}
\begin{equation}
\mathcal{S} = -\alpha_g \int d^4x \sqrt{-g} \, C_{\mu\nu\rho\sigma} C^{\mu\nu\rho\sigma},\label{action}
\end{equation}
where $\alpha_g$ represents a dimensionless coupling constant that characterizes the strength of the conformal gravitational interaction. This action is conformally invariant specifically in four spacetime dimensions, since the conformal weight of $C_{\mu\nu\rho\sigma}C^{\mu\nu\rho\sigma}$ (weight $-4$) is exactly compensated by $\sqrt{-g} \to \Omega^4\sqrt{-g}$ only for $D = 4$; in general $D$ dimensions, $\sqrt{-g}\to\Omega^D\sqrt{-g}$ while $C^2$ transforms with weight $-4$ regardless of $D$, so the cancellation fails for $D \neq 4$~\cite{isrplyxx1,isrplyxx3}. The conformal Weyl tensor in four spacetime dimensions, originally formulated by Weyl and later developed by Bach \cite{Weyl:1918pdp,Bach:1921zdq}, is explicitly expressed as
\begin{equation}
C_{\mu\nu\rho\sigma}=R_{\mu\nu\rho\sigma}-2 R_{[\mu[\rho} g_{\sigma]\nu]} +\frac{1}{3} g_{\mu[\rho}g_{\sigma]\nu} R,
\end{equation}
where $R_{\mu\nu\rho\sigma}$ and $R_{\mu\nu}$ denote the Riemann and Ricci tensors respectively, while $R$ represents the scalar curvature. The conformally covariant Weyl tensor inherits the same fundamental symmetry properties as the ordinary Riemann tensor, satisfying the essential relations
\begin{equation}
C_{\mu\nu\rho\sigma}=C_{[\mu\nu][\rho\sigma]}=C_{\rho\sigma\mu\nu},
\end{equation}
\begin{equation}
C_{\mu[\nu\rho\sigma]}=0, \qquad C^\mu{_{\nu\mu\sigma}}=0.
\end{equation}
It is noteworthy that this tensor identically vanishes in $2+1$-dimensional spacetimes, where it is replaced by the Cotton tensor, highlighting the special role of four-dimensional geometry in conformal gravity theories \cite{Riegert:1984zz}.

The explicit form of the Weyl square term appearing in the action in Eq.~\eqref{action} can be decomposed as\footnote{By utilizing the topological Gauss-Bonnet invariant in conjunction with Eq.~\eqref{explicitcsquare}, the primary action in Eq.~\eqref{action} admits an alternative representation:
\begin{equation*}
\mathcal{S} = -2\alpha_g \int d^4x \sqrt{-g} \, (R^2_{\mu\nu}-\tfrac{1}{3}R^2).
\end{equation*}}
\begin{equation}
C_{\mu\nu\rho\sigma} C^{\mu\nu\rho\sigma}=R_{\mu\nu\rho\sigma} R^{\mu\nu\rho\sigma}-2 R_{\mu\nu}R^{\mu\nu}+\frac{1}{3}R^2.\label{explicitcsquare}
\end{equation}

A crucial distinguishing feature of CWG compared to Einstein-Hilbert gravity lies in its respect for local conformal symmetry. The theory remains invariant under pointwise rescalings of the metric tensor according to $g_{\mu\nu} \rightarrow \Omega^2(x) g_{\mu\nu}$, where $\Omega(x)$ represents an arbitrary point-dependent positive function\footnote{Under such local conformal transformations, the Weyl tensor transforms as $C_{\mu\nu\rho\sigma} \rightarrow C_{\mu\nu\rho\sigma}^{'}=\Omega^2(x) C_{\mu\nu\rho\sigma}$, which constitutes conformal \emph{covariance} (weight $+2$), not invariance; only the mixed-index form $C^{\mu}{}_{\nu\rho\sigma}$ is conformally invariant. The Weyl-squared scalar $C_{\mu\nu\rho\sigma}C^{\mu\nu\rho\sigma}$ transforms with weight $\Omega^{-4}$, which is compensated by $\sqrt{-g}\to\Omega^{4}\sqrt{-g}$ in four dimensions, preserving the form of the action.}. This enhanced symmetry structure has profound implications for the resulting field equations and vacuum solutions \cite{Mannheim:1988dj}.

The field equations governing the dynamics of matter fields coupled to CWG take the form
\begin{equation}
\frac{1}{\sqrt{-g}}\, \frac{\delta \mathcal{S}}{\delta g_{\mu\nu}}=-2\alpha_{g} B^{\mu\nu}=-\frac{1}{2} T^{\mu\nu},
\end{equation}
where $T_{\mu\nu}$ represents the energy-momentum tensor of matter fields, and $B^{\mu\nu}$ denotes the celebrated Bach tensor, defined through the expression
\begin{equation}
B^{\mu\nu} =\nabla_\alpha \nabla_\beta C^{\mu\alpha\nu\beta}-\frac{1}{2}R^{\alpha\beta} C_{\mu\alpha\nu\beta}.
\label{Bachtensorexpl}
\end{equation}
This tensor possesses remarkable mathematical properties: it is symmetric, traceless, and divergence-free, making it the natural generalization of the Einstein tensor to fourth-order gravity theories. Within the CWG framework, Birkhoff's theorem continues to hold, and the cosmological constant emerges naturally as an integration constant within the vacuum solutions of the model \cite{stelle1978classical}.

Due to its gauge-like local symmetry structure, the CWG does not inherently define any characteristic length or mass scale. The spontaneous breaking of local conformal symmetry generates the dimensionful Newton's constant through a mechanism analogous to the Higgs mechanism in particle physics \cite{Mannheim:1999nc}\footnote{Alternative realizations of local Weyl symmetry can be achieved through the introduction of additional abelian gauge and scalar fields within Weyl's \emph{geometric} framework. For comprehensive treatments, see \cite{Tanhayi:2012nn,Dengiz:2024hms} for Weyl gauge-invariant quadratic and cubic gravity models, and \cite{tHooft:2014swy, Ghilencea:2021lpa, Pawlowski:1994kq,Dengiz:2016eoo} for the integration of local conformal symmetry with Standard Model gauge symmetries, and \cite{DengizWTMG1,DengizWTMG2} (and references therein) for the local Weyl conformal invariant Topologically Massive Gravity.}. Recent investigations into the fundamental features and phenomenological consequences of CWG have been extensively documented in the literature \cite{Sultana:2012qp, Konoplya:2025mvj, Jizba:2024owd,Yue:2025lkm,Yue:2025tkw,Harada:2023rqw,Burikham:2023bil,Lobo:2008zu,Said:2012xt}, revealing its rich mathematical structure and potential astrophysical applications.

A fundamental observation is that in vacuum regions where the Ricci tensor vanishes, the Bach tensor $B^{\mu\nu}$ also vanishes identically. Consequently, all vacuum solutions of Einstein's field equations automatically satisfy the CWG field equations, though the converse is not generally true, indicating that CWG admits a broader class of vacuum configurations than GR \cite{Mannheim:1988dj}\footnote{This can be easily seen as one recasts Eq.~\eqref{Bachtensorexpl} into the form with Schouten tensor $\mathcal{P}_{\mu\nu}$ as $B_{\mu\nu}=2\nabla^\alpha \nabla_{[\alpha} \mathcal{P}_{\mu]\nu}-\mathcal{P}^{\alpha\beta} C_{\mu \alpha \nu \beta}$, with $\mathcal{P}_{\mu\nu}=(1/2)(R_{\mu\nu}-(1/6)g_{\mu\nu}R)$ \cite{Besse:1987pua}. Since $\mathcal{P}_{\mu\nu}=(1/6)\Lambda g_{\mu\nu}$ in Einstein spaces (with $R_{\mu\nu}=\Lambda g_{\mu\nu}$ and $R=4\Lambda$), all the solutions of Einstein spaces are also those of CWE albeit its invalid reverse. For an informative study, see also \cite{Jizba:2024owd}.}

For static, spherically symmetric spacetimes, the most general metric ansatz takes the standard form
\begin{equation}
ds^2 = -A(r) dt^2 + B(r) dr^2 + r^2 (d\theta^2 + \sin^2 \theta d\phi^2),
\end{equation}
where the functions $A(r)$ and $B(r)$ must be determined by solving the appropriate field equations derived from the CWG action.

Through systematic analysis of static, spherically symmetric vacuum solutions to the fourth-order Bach field equations, Mannheim and Kazanas discovered an exact analytical solution that we designate as the CWGBH metric. The temporal and radial metric components are governed by the remarkable function \cite{Mannheim:1988dj,Mannheim:2011ds,Mannheim:1999bu,Mannheim:1989jh}
\begin{equation}
A(r)={B(r)}^{-1} = 1 - 3\beta\gamma - \frac{\beta(2 - 3\beta\gamma)}{r} + \gamma r + k r^2.
\label{105}
\end{equation}
This solution exhibits dramatic departures from the classical Schwarzschild BH geometry of GR. The appearance of both a linear potential term $\gamma r$ and a cosmological term $k r^2$ arises directly from the fourth-order nature of the conformal field equations, representing genuine quantum gravitational corrections absent in Einstein's theory. The parameter $\beta$ functions analogously to the gravitational mass parameter in GR, while $\gamma$ introduces scale-dependent modifications to the gravitational potential, and $k$ behaves as an effective cosmological constant \cite{Mannheim:1988dj}.

The integration constant $\gamma$ has received considerable attention in astrophysical contexts, where it has been interpreted as quantifying large-scale deviations from Newtonian gravity. This parameter plays a crucial role in modeling galactic rotation curves without requiring the introduction of dark matter, offering an alternative explanation for observed galactic dynamics. The classical Schwarzschild solution is recovered as a special limiting case when $\gamma \to 0$, $k \to 0$, and $\beta = 2GM$.

The radial structure of the metric function reveals distinct physical regimes: the term $-\beta(2 - 3\beta\gamma)/r$ dominates the Newtonian gravitational potential at short distances, the linear term $\gamma r$ becomes significant at intermediate scales, while the quadratic term $k r^2$ governs the behavior at cosmological distances. The zeros of $B(r)$ determine the locations of event horizons, and depending on the relative magnitudes and signs of the parameters $\beta$, $\gamma$, and $k$, the solution can exhibit multiple real, positive roots corresponding to distinct BH horizons.

From a geometric perspective, the MK or CWGBH metric represents a non-trivial vacuum solution of the fourth-order Bach equations, which replace Einstein's second-order field equations in CWG. These higher-order equations admit more general vacuum configurations due to their enhanced derivative structure and enlarged symmetry content. The associated curvature invariants reveal the presence of curvature singularities at $r = 0$, maintaining consistency with classical BH behavior, while the causal structure defined by the horizon locations provides a rich framework for studying thermodynamic properties.

\section{Hawking Radiation of CWGBHs via Hamilton-Jacobi Formalism of Tunneling Particles}\label{isec3}

The investigation of quantum tunneling processes for scalar particles in curved spacetime geometries provides a powerful framework for understanding thermal radiation from BHs. The relativistic Hamilton-Jacobi equation offers an elegant and computationally tractable approach to analyze such phenomena within the context of CWG. Given the spacetime geometry characterized by the MK solution in Eq.~\eqref{105}, we examine the behavior of a scalar field described by an action $S$ that satisfies the fundamental constraint \cite{srinivasan1999particle}
\begin{equation}
g^{\mu\nu} \partial_\mu S \partial_\nu S + m^2 = 0.
\end{equation}
The inverse components of the metric tensor for the CWGBH spacetime are explicitly given by
\begin{align}
g^{tt}&= -B(r), \quad g^{rr} = \frac{1}{B(r)}, \notag \\
g^{\theta\theta} &= \frac{1}{r^2}, \quad g^{\phi\phi} = \frac{1}{r^2 \sin^2\theta},
\end{align}
where, as mentioned above, $B(r)$ represents the radial metric function derived from the CWG field equations. Substituting these metric components into the Hamilton-Jacobi equation yields the explicit form
\begin{equation}
-\frac{(\partial_t S)^2}{B(r)} + B(r)(\partial_r S)^2 + \frac{(\partial_\theta S)^2}{r^2} + \frac{(\partial_\phi S)^2}{r^2 \sin^2 \theta} + m^2 = 0.
\end{equation}

For the analysis of radial tunneling processes, we focus attention on purely radial trajectories of outgoing particles and employ the standard separable ansatz for the action
\begin{equation}
S = -E t + W(r) + J_\theta\theta + J_\phi \phi,
\end{equation}
where $E$ represents the energy of the tunneling particle, $W(r)$ denotes the radial component of the action, and $J_\theta$, $J_\phi$ are the angular momentum quantum numbers. For purely radial motion, we set the angular momenta to zero ($J_\theta = J_\phi = 0$), which simplifies the analysis considerably and allows us to focus on the essential physics of the tunneling process.

The temporal and radial derivatives of the action are then given by
\begin{equation}
\partial_t S = -E, \quad \partial_r S = \frac{dW}{dr}.
\end{equation}
Substituting these expressions into the Hamilton-Jacobi equation and solving for the radial derivative yields
\begin{equation}
-\frac{E^2}{B(r)} + B(r) \left(\frac{dW}{dr}\right)^2 + m^2 = 0,
\end{equation}
which can be rearranged to obtain
\begin{equation}
\frac{dW}{dr} = \frac{\sqrt{E^2 - B(r) m^2}}{B(r)}.
\end{equation}
The radial contribution to the total action is therefore expressed as the integral
\begin{equation}
W(r) = \int \frac{\sqrt{E^2 - B(r) m^2}}{B(r)} dr.
\end{equation}
Near the event horizon located at $r = r_h$, where the metric function vanishes identically ($B(r_h) = 0$), the integrand develops a singular behavior that requires careful mathematical treatment. To handle this singularity appropriately, we introduce a proper radial coordinate transformation defined by
\begin{equation}
d\xi = \frac{dr}{\sqrt{B(r)}}.
\end{equation}
Expanding the metric function $B(r)$ in a Taylor series near the horizon yields $B(r) \approx B'(r_h)(r - r_h)$ for small deviations from the horizon radius. Consequently, the proper radial distance behaves as
\begin{equation}
\xi \approx \frac{2 \sqrt{r - r_h}}{\sqrt{B'(r_h)}}.
\end{equation}
Expressing the action integral in terms of this new coordinate system, we obtain
\begin{equation}
W(\xi) = \int \frac{\sqrt{E^2 - B(r(\xi)) m^2}}{\sqrt{B(r(\xi))}} d\xi.
\end{equation}
For small values of $\xi$ in the near-horizon region, the metric function can be approximated as $B(r(\xi)) \approx \frac{B'(r_h) \xi^2}{4}$. Focusing on the physically relevant case of massless particles ($m = 0$), which dominates the high-energy Hawking radiation spectrum, the integral simplifies dramatically to
\begin{equation}
W(\xi) \approx \frac{2E}{\sqrt{B'(r_h)}} \int \frac{d\xi}{\xi}.
\end{equation}
This integral exhibits a characteristic logarithmic divergence at the horizon, which gives rise to an imaginary contribution to the action. This imaginary part is crucial for the tunneling interpretation and is given by \cite{sucu2025nonlinear}
\begin{equation}
\operatorname{Im} W = \frac{2\pi E}{B'(r_h)}.
\end{equation}
Therefore, the imaginary part of the total classical action becomes
\begin{equation}
\operatorname{Im} S = \frac{2\pi E}{B'(r_h)}.
\end{equation}
According to the semiclassical tunneling framework developed for BH radiation, the emission probability for particles to tunnel through the gravitational potential barrier is governed by \cite{sucu2023gup}
\begin{equation}
\Gamma \sim \exp\left(-\frac{4\pi E}{B'(r_h)}\right).
\end{equation}
This exponential suppression factor directly reveals the thermal nature of the radiation and allows us to identify the Hawking temperature as
\begin{equation}
T_H = \frac{B'(r_h)}{4\pi} = -\frac{3 \beta^{2} \gamma}{4 \pi r_h^{2}}+\frac{\beta}{2 \pi r_h^{2}}+\frac{\gamma}{4 \pi}+\frac{k r_h}{2 \pi}.
\label{hawking_temp}
\end{equation}
This expression represents a significant generalization of the standard Schwarzschild result and demonstrates the rich thermal structure emerging from CWG. The temperature formula is consistent with results obtained through alternative methods such as the null geodesic approach, providing an important cross-validation of the thermal spectrum derivation. It is important to note that for physically consistent BH solutions, the horizon radius $r_h$ must be determined self-consistently from the horizon condition $B(r_h) = 0$, which couples $r_h$ to the MK parameters $(\beta, \gamma, k)$. For a comprehensive analysis of the horizon structure in the MK metric, see Ref.~\cite{Turner:2020cni}.

\begin{table*}[ht!]
\centering
\renewcommand{\arraystretch}{1.6}
\setlength{\tabcolsep}{4pt}
\resizebox{\textwidth}{!}{%
\begin{tabular}{|c|c|c|c|c||c|c|c|c|c||c|c|c|c|c|}
\hline
\multicolumn{5}{|c||}{$\beta = 0.5$} & \multicolumn{5}{c||}{$\beta = 1.0$} & \multicolumn{5}{c|}{$\beta = 2.0$} \\
\hline
$k$ & $\gamma$ & $r_h$ & $T_H$ & $\Delta T_H$ & $k$ & $\gamma$ & $r_h$ & $T_H$ & $\Delta T_H$ & $k$ & $\gamma$ & $r_h$ & $T_H$ & $\Delta T_H$ \\
\hline
0 & 0 & 1.000 & 0.0796 & -- & 0 & 0 & 2.000 & 0.0398 & -- & 0 & 0 & 4.000 & 0.0199 & -- \\
0 & 0.05 & 0.988 & 0.0825 & +3.6\% & 0 & 0.05 & 1.952 & 0.0426 & +7.1\% & 0 & 0.05 & 3.817 & 0.0226 & +13.4\% \\
0 & 0.10 & 0.976 & 0.0852 & +7.1\% & 0 & 0.10 & 1.908 & 0.0451 & +13.4\% & 0 & 0.10 & 3.657 & 0.0246 & +23.8\% \\
\hline
0.05 & 0 & 0.956 & 0.0946 & +18.9\% & 0.05 & 0 & 1.738 & 0.0665 & +67.2\% & 0.05 & 0 & 2.847 & 0.0619 & +211.3\% \\
0.10 & 0 & 0.922 & 0.1083 & +36.1\% & 0.10 & 0 & 1.595 & 0.0880 & +121.1\% & 0.10 & 0 & 2.478 & 0.0913 & +358.8\% \\
\hline
0.05 & 0.05 & 0.946 & 0.0970 & +21.9\% & 0.05 & 0.05 & 1.710 & 0.0679 & +70.7\% & 0.05 & 0.05 & 2.777 & 0.0612 & +207.4\% \\
0.10 & 0.10 & 0.905 & 0.1123 & +41.1\% & 0.10 & 0.10 & 1.551 & 0.0889 & +123.3\% & 0.10 & 0.10 & 2.354 & 0.0856 & +330.4\% \\
\hline
\end{tabular}%
}
\caption{Hawking temperature $T_H$ for CWGBHs organized by mass parameter $\beta$. The horizon radii $r_h$ are self-consistently determined from $B(r_h)=0$, ensuring physically realizable BH configurations. The baseline case for each $\beta$ corresponds to the Schwarzschild limit ($\gamma=0$, $k=0$), where $r_h = 2\beta$. The percentage deviation $\Delta T_H$ quantifies the thermal enhancement relative to Schwarzschild at the same $\beta$.}
\label{tab:TH_variation}
\end{table*}

Table~\ref{tab:TH_variation} presents a comprehensive analysis of the Hawking temperature $T_H$ across different parameter regimes, with horizon radii $r_h$ determined self-consistently from the condition $B(r_h)=0$. The table is organized by the mass-like parameter $\beta$, with the Schwarzschild limit ($\gamma=0$, $k=0$) serving as the baseline for each column. In this limit, the horizon radius reduces to $r_h = 2\beta$, recovering the standard result. The percentage deviation $\Delta T_H$ quantifies the thermal enhancement relative to the Schwarzschild BH at the same $\beta$ value, thereby isolating the effects of the CWG parameters $\gamma$ and $k$. The data reveals several key physical insights. The parameter $\gamma$ introduces modest corrections to both $r_h$ and $T_H$. For $\gamma = 0.1$ with $k=0$, the horizon radius decreases slightly (e.g., from $r_h = 2.000$ to $r_h = 1.908$ for $\beta = 1$), while $T_H$ increases by approximately $7$--$24\%$ depending on $\beta$. This behavior reflects the linear potential contribution $\gamma r$ in the metric function, which modifies the gravitational potential at intermediate scales. The parameter $k$ produces more pronounced effects: for $k = 0.1$ with $\gamma = 0$, the temperature enhancement reaches $36$--$359\%$, with larger deviations occurring at higher $\beta$ values. The quadratic term $kr^2$ in $B(r)$ significantly contracts the horizon radius (e.g., from $r_h = 4.000$ to $r_h = 2.478$ for $\beta = 2$), leading to substantially higher Hawking temperatures. When both $\gamma$ and $k$ are nonzero, their effects combine to further modify the thermal properties: the combined case ($k=0.1$, $\gamma=0.1$) yields temperature enhancements of $41$--$330\%$, demonstrating the rich parameter space available in CWG for modifying BH thermal properties while maintaining physical consistency through the horizon condition.

\section{GUP Corrected Hawking Radiation of CWGBHs}\label{isec4}

Before proceeding with the GUP analysis, we address a subtle but important theoretical point concerning the compatibility of the Generalized Uncertainty Principle with Conformal Weyl Gravity. The CWG action is constructed from the Weyl tensor squared and possesses local conformal symmetry, which forbids dimensionful coupling constants such as Newton's constant $G$ in the fundamental Lagrangian. Since the GUP introduces a minimum measurable length scale $\Delta x_{\min} \sim \sqrt{\lambda}\, \ell_P$, where $\ell_P = \sqrt{\hbar G/c^3}$ is the Planck length, one might question whether these frameworks are mutually consistent.

The resolution lies in the well-established mechanism of \textit{spontaneous symmetry breaking}. As demonstrated by Mannheim and Kazanas~\cite{Mannheim:1988dj,Mannheim:1993rd}, when conformally coupled scalar fields acquire vacuum expectation values, conformal symmetry is spontaneously broken, and all necessary mass scales---including an effective Newton's constant and the Planck mass---emerge dynamically. This mechanism is entirely analogous to the Brout-Englert-Higgs mechanism in gauge theories and has been further developed in various contexts: Ghilencea~\cite{Ghilencea:2018dqd} showed that the Stueckelberg mechanism in Weyl quadratic gravity generates the Planck scale while reducing Weyl geometry to Riemannian geometry at low energies; 't~Hooft~\cite{tHooft:2014swy} proposed that local conformal symmetry is a fundamental gauge symmetry of nature that is spontaneously broken through dilaton dynamics; and Salvio and Strumia~\cite{Salvio:2014soa} demonstrated in their ``Agravity'' framework that the Planck scale emerges via dimensional transmutation from purely dimensionless couplings.

Within this paradigm, the GUP should be understood as an \textit{effective phenomenological description} of quantum gravity corrections, valid in the low-energy regime where conformal symmetry is already broken and the Planck scale has emerged. This interpretation is supported by derivations of GUP from UV-complete theories such as string theory~\cite{Konishi:1989wk} and its connection to the running Newton's constant in asymptotically safe gravity~\cite{Scardigli:2022jtt}. Our approach therefore employs CWG to provide the classical gravitational background (the MK BH solution), while GUP captures the leading-order quantum corrections to matter fields propagating on this background. This effective field theory perspective ensures complete theoretical consistency: conformal symmetry governs the UV structure of the theory, while GUP-modified thermodynamics describes the IR physics where symmetry breaking has generated the relevant scales.

To account for the profound influence of quantum gravitational effects on BH thermodynamics, one of the most promising and well-developed theoretical approaches involves the implementation of the GUP. This principle introduces a fundamental minimal length scale that emerges naturally from several candidate theories of quantum gravity, including string theory, loop quantum gravity, and non-commutative geometry \cite{gangopadhyay2015constraints,tawfik2013impacts,nozari2006comparison}. The GUP framework provides a phenomenological window into the quantum gravitational regime by modifying the standard Heisenberg uncertainty relations through the incorporation of Planck-scale corrections.

The modified uncertainty relation in its most commonly employed form is expressed as \cite{tawfik2014generalized,maggiore1993algebraic,bosso202330}
\begin{equation}
\Delta x \Delta p \geq \frac{\hbar}{2} \left(1 + \lambda^2 l_p^2 \frac{(\Delta p)^2}{\hbar^2}\right),
\end{equation}
where $\lambda$ represents a model-dependent dimensionless constant typically of order unity, and $l_p = \sqrt{\hbar G/c^3}$ denotes the fundamental Planck length scale. This deformation of the uncertainty principle can be inverted to express the momentum uncertainty in terms of the position uncertainty, yielding
\begin{equation}
\Delta p \geq \frac{\hbar}{2 \Delta x} \left(1 + \frac{2\lambda^2 l_p^2}{\Delta x^2} \right).
\end{equation}

In the context of BH physics, particularly near the event horizon where quantum effects become most pronounced, it is physically reasonable to associate the fundamental position uncertainty with the characteristic horizon length scale, leading to the identification $\Delta x \sim 2 r_h$. This association reflects the fact that the horizon represents the natural boundary beyond which classical notions of spacetime geometry begin to break down \cite{carr2011generalized,perivolaropoulos2017cosmological}.

The energy of particles emitted through quantum tunneling processes, as estimated from the uncertainty principle applied to momentum fluctuations, receives quantum gravitational corrections according to
\begin{equation}
E_{\text{GUP}} \sim \Delta p \sim \frac{\hbar c}{4 r_h} \left(1 + \frac{\lambda^2 l_p^2}{2 r_h^2} \right).
\end{equation}
This fundamental modification in the effective energy of tunneling particles directly impacts the semiclassical tunneling probability that governs Hawking radiation emission. In the standard semiclassical framework, the emission probability follows the characteristic exponential suppression \cite{gecim2017gupp}
\begin{equation}
\Gamma \sim \exp\left(-\frac{2\pi E}{\kappa}\right),
\end{equation}
where $\kappa = \frac{1}{2} B'(r_h)$ represents the surface gravity evaluated at the horizon. Under the influence of GUP corrections, this tunneling probability becomes modified to
\begin{align}
\Gamma_{\text{GUP}} &\sim \exp\left(-\frac{2\pi E_{\text{GUP}}}{\kappa}\right) \notag \\
&= \exp\left( -\frac{4\pi E}{B'(r_h)}  \left(1 + \frac{\lambda^2 l_p^2}{2 r_h^2} \right) \right).
\end{align}
From this modified exponential structure, we can directly extract the QC Hawking temperature by identifying the effective thermal factor in the Boltzmann-like distribution. This yields the GUP-corrected temperature as
\begin{align}
T_{\text{GUP}} &= \frac{B'(r_h)}{4\pi} \left(1 + \frac{\lambda^2 l_p^2}{2 r_h^2} \right)^{-1} \notag \\
&= T_H \left(1 + \frac{\lambda^2 l_p^2}{2 r_h^2} \right)^{-1},
\end{align}
where $T_H$ represents the classical Hawking temperature derived in the previous section.

\begin{figure}[ht!]
\centering
\includegraphics[width=0.95\columnwidth]{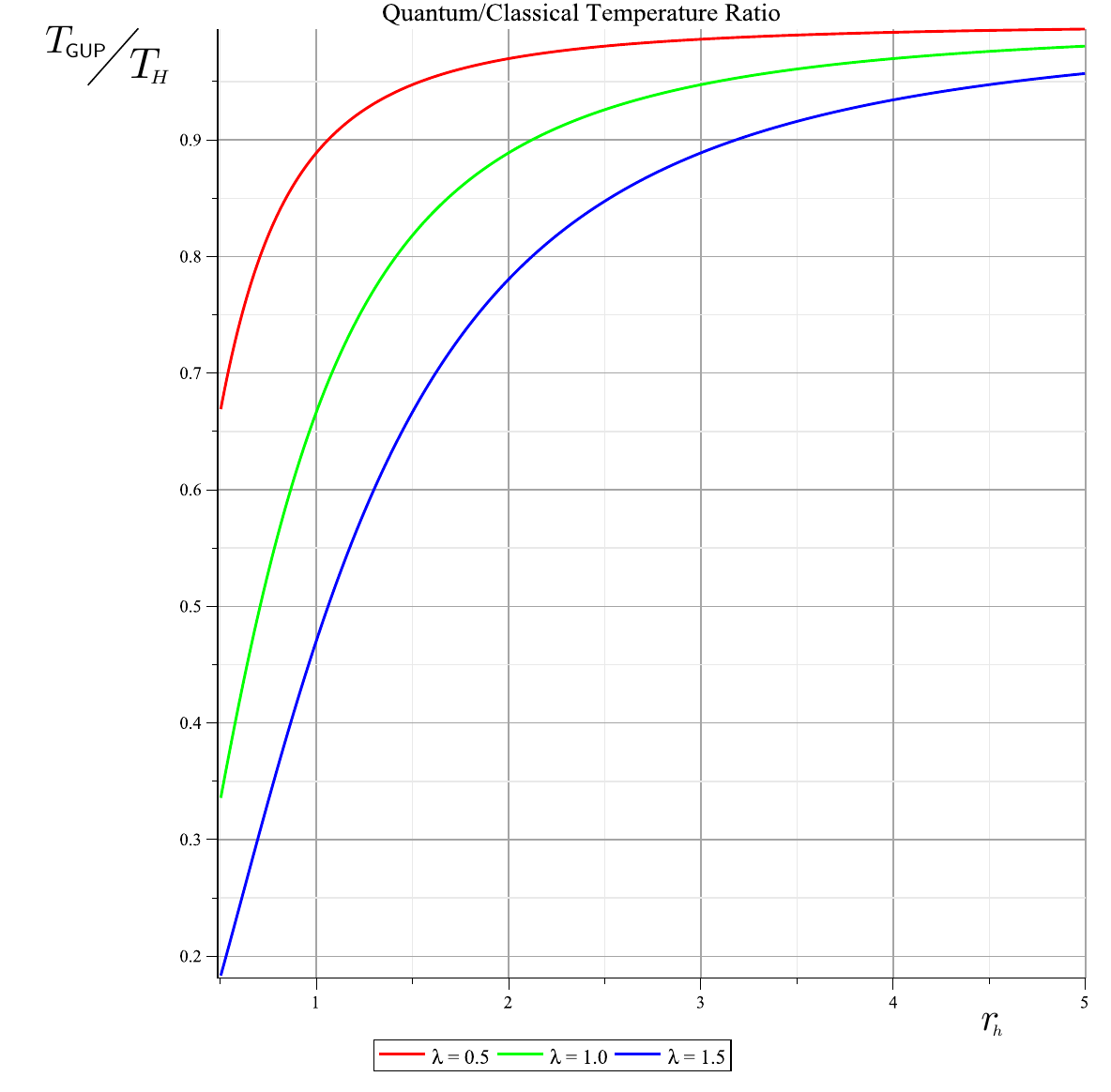}
\caption{Quantum temperature suppression factor $\Delta T/T_{\text{classical}}$ versus horizon radius $r_h$ for GUP parameters $\lambda = \{0.5, 1.0, 1.5\}$, calculated with CWG parameters $\beta = 2$, $\gamma = 1$, and $k = 1$. The suppression factor quantifies the fractional reduction in thermal radiation due to minimal length effects, exhibiting enhanced quantum corrections for smaller horizon radii. This computational analysis demonstrates the systematic energy scale modifications induced by the generalized uncertainty principle, with the suppression becoming dominant in regimes where $r_h \sim l_p$, establishing clear boundaries between semiclassical and quantum gravitational domains.}
\label{fig:TGUPsuppression}
\end{figure}

This result demonstrates that the emission spectrum undergoes a systematic redshift due to quantum gravitational corrections, with the magnitude of the effect being inversely related to the square of the horizon radius. As illustrated in Figure~\ref{fig:TGUPsuppression}, the quantum suppression factor exhibits characteristic scaling behavior that becomes increasingly pronounced for smaller horizon radii, quantitatively validating the theoretical predictions of minimal length effects in gravitational thermodynamics.

The physical interpretation of this correction is particularly illuminating: as the BH horizon radius approaches the Planck scale ($r_h \to l_p$), the quantum gravitational corrections become increasingly pronounced, effectively suppressing the radiation temperature compared to its semiclassical counterpart. While these effects remain negligible for macroscopic astrophysical BHs with horizon radii many orders of magnitude larger than the Planck length, they become critically important when investigating microscopic BHs, primordial BH evolution, or the final stages of BH evaporation where the horizon approaches Planck-scale dimensions. The GUP framework thus provides an essential conceptual and computational bridge between standard semiclassical thermodynamic descriptions and the quantum gravitational regimes where conventional field theory breaks down.

The phase transition implications of GUP corrections become manifest through the modified critical behavior, fundamentally altering the thermal energy scales at which thermodynamic instabilities emerge \cite{Adler:2001vs,Nozari:2008rc,Tawfik:2014gra}. For CWGBHs, the quantum-corrected critical temperature exhibits the form
\begin{equation}
T_{\text{crit,GUP}} = T_{\text{crit}} \left(1 + \frac{\lambda^2 l_p^2}{2r_h^2}\right)^{-1},
\end{equation}
where $T_{\text{crit}}$ represents the classical critical temperature determined by the heat capacity divergence condition $kr_h^3 + 3\beta^2\gamma - 2\beta = 0$ at the critical horizon radius $r_{\text{crit}}$. The explicit expression for the classical critical temperature follows from the surface gravity evaluation
\begin{align}
T_{\text{crit}} &= \frac{B'(r_{\text{crit}})}{4\pi} \notag \\
&= \frac{3\beta^2\gamma}{4\pi r_{\text{crit}}^2} + \frac{\beta}{2\pi r_{\text{crit}}^2} + \frac{\gamma}{4\pi} + \frac{kr_{\text{crit}}}{2\pi},
\end{align}
characterizing the thermal energy boundary separating thermodynamically stable and unstable phases in the absence of quantum gravitational corrections.

\begin{figure}[ht!]
\centering
\includegraphics[width=0.95\columnwidth]{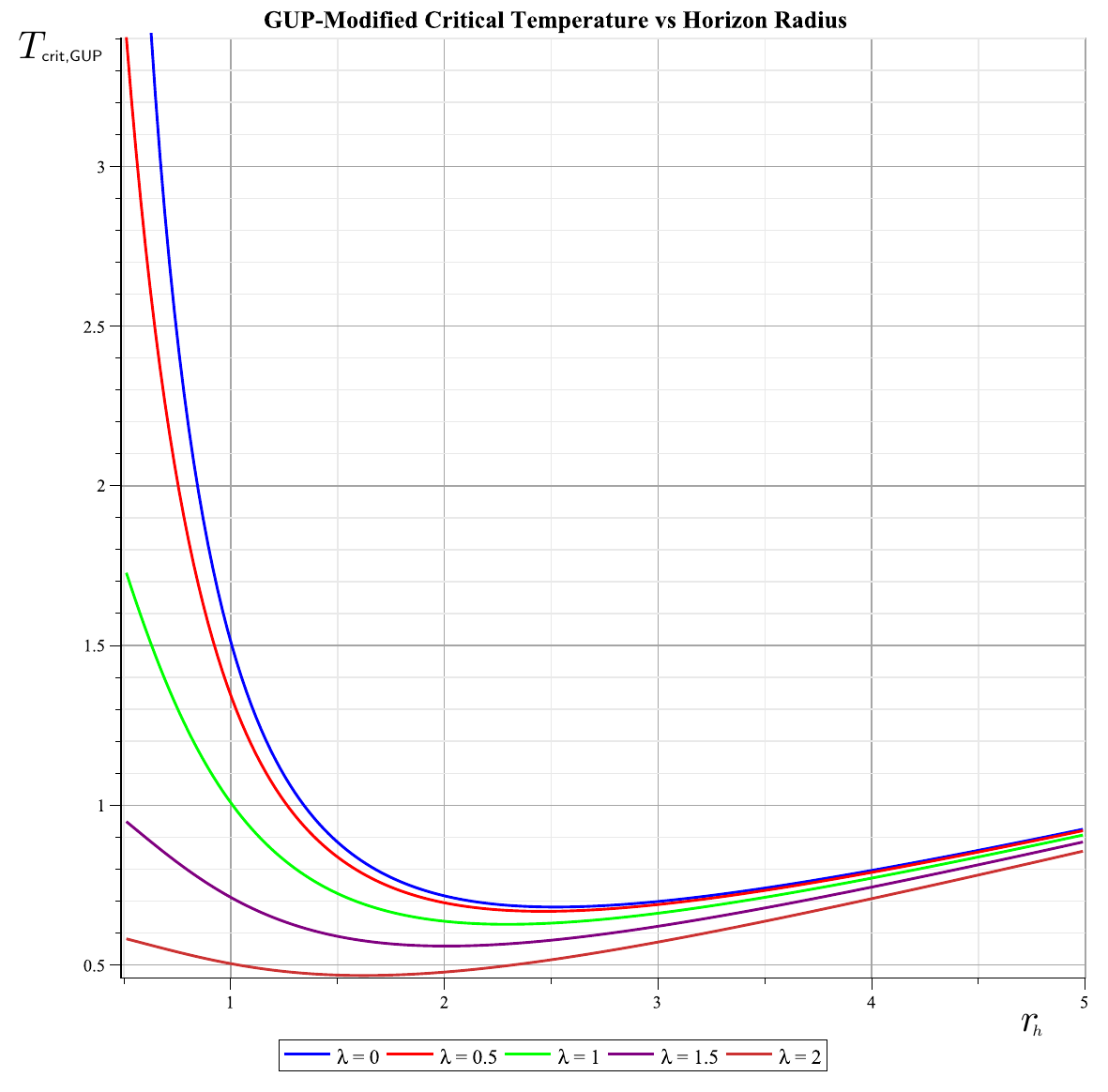}
\caption{GUP-modified critical temperature as a function of event horizon radius $r_h$ for various quantum correction parameters $\lambda = \{0, 0.5, 1.0, 1.5, 2.0\}$, with fixed CWG parameters $\beta = 2$, $\gamma = 1$, and $k = 1$. The classical limit ($\lambda = 0$) exhibits the standard inverse-square scaling behavior, while increasing GUP parameters systematically suppress the critical temperature across all horizon scales. The quantum suppression becomes increasingly pronounced for smaller horizon radii, demonstrating the enhanced significance of minimal length effects in the near-Planckian regime where $r_h \to l_p$.}
\label{fig:TGUPcrit}
\end{figure}

The GUP parameter $\lambda$ introduces a characteristic quantum length scale that systematically suppresses the critical temperature through minimal length effects, thereby modifying phase boundaries and shifting the stability regions identified through heat capacity analysis toward lower energy scales \cite{Pedram:2011gj,Gecim:2019pft,Pourhassan:2020pv}. As demonstrated in Figure~\ref{fig:TGUPcrit}, the quantum-corrected critical temperature exhibits systematic suppression that becomes increasingly pronounced for smaller horizon radii, reflecting the enhanced significance of minimal length corrections in the near-Planckian regime.

These quantum corrections demonstrate systematic alterations to thermodynamic phase structures in the near-Planckian regime, where the quantum suppression factor becomes increasingly pronounced as $r_h \to l_p$, establishing domains where conventional semiclassical descriptions become inadequate for capturing the full spectrum of gravitational phenomena \cite{Cai:2009ua,Hossenfelder:2013eka}.

\section{QC Thermodynamics of CWGBHs}\label{isec5}

In the regime where BH dimensions approach the Planck scale, classical thermodynamic descriptions become fundamentally inadequate, necessitating the incorporation of quantum corrections that capture the underlying microscopic degrees of freedom. One of the most sensitive thermodynamic quantities to such quantum effects is the entropy, which in the classical Bekenstein-Hawking framework is simply proportional to the event horizon area. However, this classical description fails to account for the rich quantum structure that emerges near the Planck scale \cite{cognola1995one}.

To provide a more complete theoretical framework, quantum statistical mechanics offers a sophisticated approach through systematic counting of horizon microstates \cite{ghosh2006counting,krasnov1998quantum}. In canonical ensembles with fixed energy and particle number, quantum fluctuations give rise to subleading modifications to the classical entropy expressions. These corrections effectively capture the statistical uncertainties arising from quantum degrees of freedom that are entirely absent in semiclassical treatments, providing crucial insights into the quantum nature of gravitational thermodynamics \cite{chui1992temperature,gursel2025thermodynamics,sucu2025astrophysical}.

One particularly well-motivated proposal involves the exponentially corrected entropy form, which naturally emerges from specific microstate counting schemes in various approaches to quantum gravity \cite{chatterjee2020exponential,sucu2025exploring}.
\begin{equation}
S = S_0 + e^{-S_0}, \label{expcorrected}
\end{equation}
where $S_0$ represents the classical Bekenstein-Hawking entropy. The exponential correction term becomes increasingly relevant for small values of $S_0$, which corresponds precisely to near-extremal or Planck-scale BHs where quantum effects are expected to dominate.

We note that alternative quantum corrections to BH entropy have been proposed in the literature, most notably the logarithmic form $S = S_0 - \frac{3}{2}\ln S_0 + \ldots$, which arises from one-loop quantum fluctuations around the saddle-point geometry in the Euclidean path integral~\cite{isrplyxx4,isrplyxx5,isrplyxx6}. While logarithmic corrections dominate perturbative quantum effects for large BHs, the exponential form $S = S_0 + e^{-S_0}$ captures non-perturbative effects associated with the discrete nature of the BH microstate spectrum~\cite{chatterjee2020exponential}. Since our analysis focuses on the near-Planckian domain where $S_0 \sim \mathcal{O}(1)$ and GUP corrections are most operative, the exponential correction provides the physically appropriate non-perturbative completion. For large BHs ($S_0 \gg 1$), the exponential term $e^{-S_0} \to 0$ and the standard Bekenstein-Hawking entropy is recovered.

Regarding the choice of $S_0 = S_{BH} = \pi r_h^2$ as the base entropy: in higher-derivative theories such as CWG, the Wald entropy~\cite{isrplyxx7,isrplyxx8} generically differs from the Bekenstein-Hawking area law. However, for the MK vacuum solution in the symmetry-broken phase---where the effective Newton constant $G_{\rm eff}$ emerges dynamically~\cite{Mannheim:1988dj,isrplyxx9}---the Wald entropy evaluated on the static, spherically symmetric horizon reduces to the standard area law $S_{\rm Wald} = \pi r_h^2$ (in units $G_{\rm eff} = 1$)~\cite{isrplyxx10}. This equivalence holds because the Weyl-squared contributions to the Wald integrand produce only area-proportional terms for Einstein space solutions on a two-sphere horizon cross-section~\cite{isrplyxx11}. Therefore, $S_0 = S_{BH} = \pi r_h^2$ is the appropriate starting point for the exponential correction scheme applied to the MK metric.

It is important to emphasize that the thermodynamic quantities presented in this section are computed as parametric functions of the horizon radius $r_h$ and the CWG parameters $(\beta, \gamma, k)$. This parametric approach is standard in BH thermodynamics and reveals the functional dependence of thermodynamic potentials on horizon geometry. For numerical calculations requiring specific horizon values, one should employ the self-consistent approach where $r_h$ is determined from $B(r_h) = 0$, as detailed in Section~\ref{isec3}. Table~\ref{tab:entropy} presents explicit numerical values of the Bekenstein-Hawking entropy $S_{BH} = \pi r_h^2$ and Hawking temperature $T_H$ computed using self-consistent horizon radii for various CWG parameter combinations.

\begin{table}[ht!]
\centering
\renewcommand{\arraystretch}{1.5}
\setlength{\tabcolsep}{4pt}
\begin{tabular}{|c|c|c|c|c|c|}
\hline
$\beta$ & $\gamma$ & $k$ & $r_h$ & $S_{BH}$ & $T_H$ \\
\hline\hline
0.5 & 0.00 & 0.00 & 1.0000 & 3.1416 & 0.079577 \\
1.0 & 0.00 & 0.00 & 2.0000 & 12.5664 & 0.039789 \\
2.0 & 0.00 & 0.00 & 4.0000 & 50.2655 & 0.019894 \\
\hline
0.5 & 0.05 & 0.00 & 0.9878 & 3.0654 & 0.082476 \\
1.0 & 0.05 & 0.00 & 1.9523 & 11.9738 & 0.042605 \\
2.0 & 0.05 & 0.00 & 3.8167 & 45.7631 & 0.022553 \\
\hline
0.5 & 0.10 & 0.00 & 0.9761 & 2.9934 & 0.085210 \\
1.0 & 0.10 & 0.00 & 1.9083 & 11.4408 & 0.045106 \\
2.0 & 0.10 & 0.00 & 3.6569 & 42.0112 & 0.024620 \\
\hline
0.5 & 0.00 & 0.05 & 0.9563 & 2.8729 & 0.094631 \\
1.0 & 0.00 & 0.05 & 1.7377 & 9.4859 & 0.066538 \\
2.0 & 0.00 & 0.05 & 2.8466 & 25.4574 & 0.061934 \\
\hline
0.5 & 0.05 & 0.05 & 0.9463 & 2.8134 & 0.097038 \\
1.0 & 0.05 & 0.05 & 1.7102 & 9.1884 & 0.067923 \\
2.0 & 0.05 & 0.05 & 2.7769 & 24.2250 & 0.061164 \\
\hline
\end{tabular}
\caption{Bekenstein-Hawking entropy $S_{BH} = \pi r_h^2$ and Hawking temperature $T_H$ for CWGBHs with self-consistent horizon radii $r_h$ determined from $B(r_h) = 0$. The first three rows correspond to the Schwarzschild limit ($\gamma = 0$, $k = 0$), where $r_h = 2\beta$ and $S_{BH} = 4\pi\beta^2$ are exactly recovered. The CWG parameters systematically reduce the horizon radius and entropy while increasing the temperature, consistent with the thermodynamic principle that smaller BHs radiate more intensely.}
\label{tab:entropy}
\end{table}

Using this entropy modification within the framework of CWG, the corrected internal energy can be systematically computed by applying the first law of BH thermodynamics
\begin{equation}
E = \int T_H \, dS, \label{s41}
\end{equation}
where $T_H$ denotes the Hawking temperature derived in Eq.~\eqref{hawking_temp}. For macroscopic BHs with $S_0 \gg 1$, the exponential correction $e^{-S_0}$ is negligible (e.g., $e^{-\pi r_h^2} \approx 4.3\times 10^{-2}$ at $r_h = 1$ and $\approx 3.5\times 10^{-6}$ at $r_h = 2$), so the integration measure reduces to $dS \approx dS_0 = 2\pi r_h\,dr_h$, with the expansion parameter being $e^{-S_0}$. The ``$\approx$'' signs in the following expressions indicate that terms of order $\mathcal{O}(e^{-\pi r_h^2})$ have been dropped. Substituting the QC entropy expression and performing the integration yields
\begin{equation}
E \approx \left(\beta - \tfrac{3}{2}\beta^2\gamma\right)\ln r_h + \frac{\gamma\,r_h^2}{4} + \frac{k\,r_h^3}{3}. \label{s42}
\end{equation}
This expression is defined up to an additive constant of integration~\cite{isrplyxx12}. In the Schwarzschild limit ($\gamma = 0$, $k = 0$), it reduces to $E = \beta\ln r_h$, which has the correct dimension $[\text{mass}]\times[\text{dimensionless}]$. The first-law consistency $dE/dr_h = T_H \cdot dS_0/dr_h$ holds identically by construction.

\begin{figure}[ht!]
    \centering
    \includegraphics[width=0.95\columnwidth]{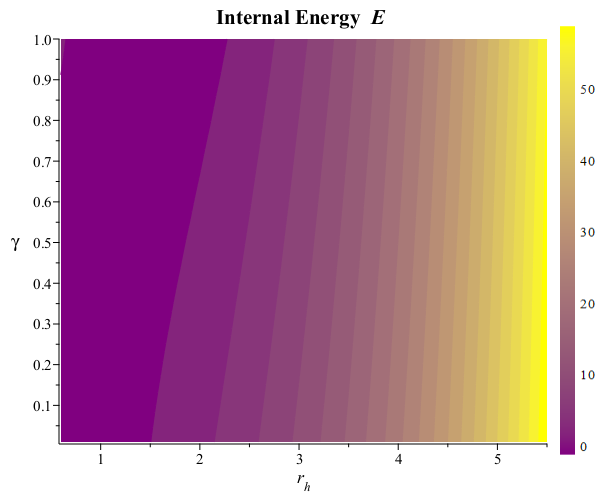}
    \caption{Filled contour plot of the internal energy $E$ as a function of the event horizon radius $r_h$ and the scale-dependent parameter $\gamma$, with fixed values $\beta = 2$ and $k = 1$. The contour lines with colorbar quantitatively demonstrate the rapid increase of energy with $r_h$, while the near-vertical contour structure confirms the minimal influence of $\gamma$ across most of the parameter space, highlighting the dominant role of horizon size in determining the BH mass-energy content.}
    \label{EEe}
\end{figure}

Figure~\ref{EEe} presents a filled contour plot illustrating the behavior of the internal energy $E$ as a function of both the event horizon radius $r_h$ and the scale-dependent parameter $\gamma$, with fixed parameter values $\beta = 2$ and $k = 1$. The visualization clearly demonstrates that the energy increases monotonically with $r_h$, consistent with the thermodynamic interpretation of BH mass. The contour lines, which appear nearly vertical across the parameter space, indicate that $E$ depends predominantly on $r_h$ with only weak sensitivity to $\gamma$. The colorbar scale ranges from approximately $0$ to $55$ units. The contour structure reveals that $\gamma$ exerts a negligible influence on $E$ at larger horizon scales, but becomes increasingly important in the small $r_h$ regime. This behavior indicates that scale-dependent geometric corrections subtly modify the BH energy structure near the Planckian regime, suggesting that quantum corrections encoded in $\gamma$ become critically relevant in the small-scale limit and may significantly influence evaporation dynamics or remnant formation scenarios.

The Helmholtz free energy, which provides crucial information about the system's equilibrium properties, is defined through the standard thermodynamic relation \cite{WOS:001565141800002NPB}
\begin{equation}
F = E - T_H S_0, \label{s43}
\end{equation}
yielding upon substitution
\begin{equation}
F \approx \left(\beta - \tfrac{3}{2}\beta^2\gamma\right)\ln r_h - \frac{k\,r_h^3}{6} + \frac{3\beta^2\gamma}{4} - \frac{\beta}{2}. \label{free}
\end{equation}

\begin{figure}[ht!]
    \centering
    \includegraphics[width=0.95\columnwidth]{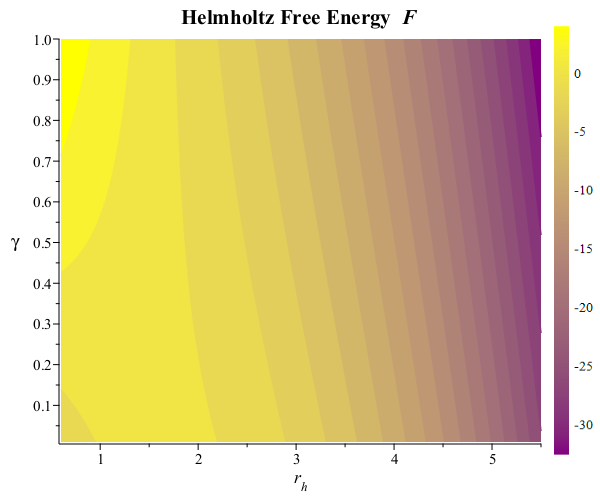}
    \caption{Filled contour plot of the Helmholtz free energy $F$ displaying the QC thermodynamic landscape for CWGBHs, with fixed parameters $\beta = 2$ and $k = 1$. The colorbar quantifies the transition from positive (yellow, small $r_h$) to negative (purple, large $r_h$) free energy values, with contour lines illustrating the complex parameter dependence that governs equilibrium stability conditions. The predominantly vertical contour structure at large $r_h$ indicates $r_h$-dominated behavior, while deviations at small $r_h$ reveal $\gamma$-dependent quantum corrections.}
    \label{FFF}
\end{figure}

Figure~\ref{FFF} displays the Helmholtz free energy behavior, revealing the complex interplay between geometric parameters and thermodynamic stability. The filled contour representation with colorbar shows $F$ taking predominantly positive values (yellow regions) at small $r_h$ and transitioning to increasingly negative values (purple regions) as $r_h$ increases, with the colorbar indicating a range from approximately $0$ to $-30$ units. The contour structure demonstrates that larger BHs possess more negative free energy, indicating enhanced thermodynamic stability. The slight curvature of contour lines at small $r_h$ reveals the region where $\gamma$-dependent corrections become non-negligible.

The QC pressure, fundamental for understanding the mechanical properties of the BH system, follows from the thermodynamic relation \cite{gursel2025thermodynamics}
\begin{equation}
P = -\frac{d F}{d V}, \label{pressure}
\end{equation}
which evaluates to the expression
\begin{equation}
P \approx \frac{k\,r_h^3 + 3\beta^2\gamma - 2\beta}{8\pi\,r_h^3}.\label{pres}
\end{equation}
In the Schwarzschild limit ($\gamma = 0$, $k = 0$), this reduces to $P = -\beta/(4\pi r_h^3)$.

\begin{figure}[ht!]
    \centering
    \includegraphics[width=0.95\columnwidth]{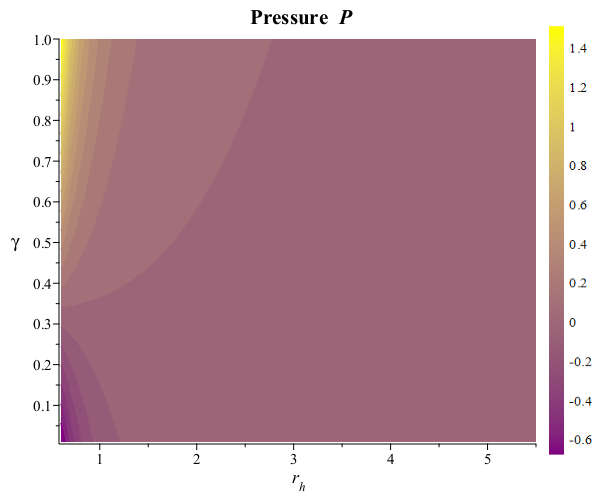}
    \caption{Filled contour plot of the QC pressure $P$ as a function of horizon parameters $r_h$ and $\gamma$, with fixed $\beta = 2$ and $k = 1$. The colorbar quantifies pressure values ranging from negative (purple) to positive (yellow), while the curved contour structure illustrates the non-trivial dependence on both parameters. The mechanical response properties of CWGBHs exhibit significantly richer structure than the energy-type potentials, with the $\gamma$ parameter playing a more pronounced role in determining pressure characteristics.}
    \label{PPP}
\end{figure}

Figure~\ref{PPP} presents the QC pressure as a filled contour plot, revealing a rich two-dimensional structure that distinguishes it from the other thermodynamic potentials. The colorbar scale ranges from approximately $-0.6$ to $1.4$ units, with the contour lines exhibiting significant curvature throughout the $(r_h, \gamma)$ parameter space. Notably, the pressure shows strong dependence on both parameters: at fixed $r_h$, increasing $\gamma$ systematically increases the pressure, while at fixed $\gamma$, the pressure grows with $r_h$. The region of negative pressure (dark purple) at small $r_h$ and small $\gamma$ corresponds to configurations where quantum corrections and geometric effects compete, potentially indicating exotic thermodynamic behavior characteristic of the CWG framework.

The enthalpy, incorporating quantum corrections and representing the total energy content under constant pressure conditions, follows as
\begin{equation}
H = E + P V, \label{s35}
\end{equation}
resulting in the expression
\begin{align}
H &\approx \left(\beta - \tfrac{3}{2}\beta^2\gamma\right)\ln r_h + \frac{\gamma\,r_h^2}{4} \notag \\
&\quad + \frac{k\,r_h^3}{2} + \frac{\beta^2\gamma}{2} - \frac{\beta}{3}. \label{s36}
\end{align}

\begin{figure}[ht!]
    \centering
    \includegraphics[width=0.95\columnwidth]{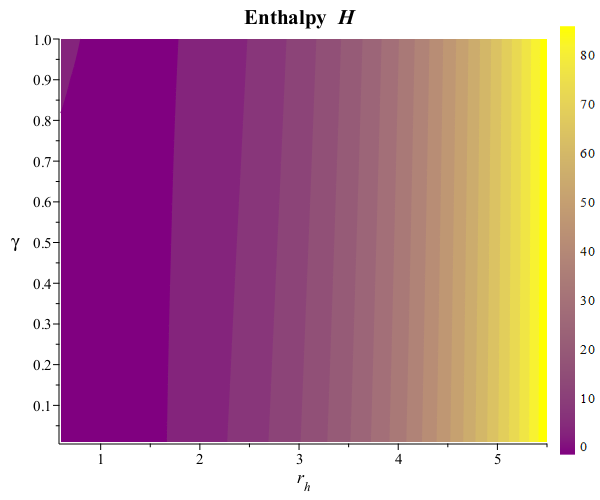}
    \caption{Filled contour plot of the enthalpy $H$ demonstrating the total energy content of CWGBHs under constant pressure conditions, with fixed parameters $\beta = 2$ and $k = 1$. The colorbar quantifies enthalpy values, while the near-vertical contour structure reveals the quantum-enhanced thermodynamic behavior dominated by horizon radius dependence.}
    \label{HHH}
\end{figure}

Figure~\ref{HHH} displays the enthalpy distribution as a filled contour plot, demonstrating behavior qualitatively similar to the internal energy $E$. The colorbar scale extends from approximately $0$ to $85$ units. The nearly vertical contour lines confirm that $H$ is predominantly determined by $r_h$, with $\gamma$ producing only minor modifications. This $r_h$-dominated behavior reflects the fact that both $E$ and $PV$ scale strongly with horizon radius, making the enthalpy an excellent proxy for the total thermodynamic energy content of the BH system.

The Gibbs free energy, essential for assessing global thermodynamic stability and phase transition behavior, becomes
\begin{equation}
G \approx -\frac{1}{12}\left(6\ln r_h - 5\right)\left(3\beta\gamma - 2\right)\beta.
\end{equation}

\begin{figure}[ht!]
    \centering
    \includegraphics[width=0.95\columnwidth]{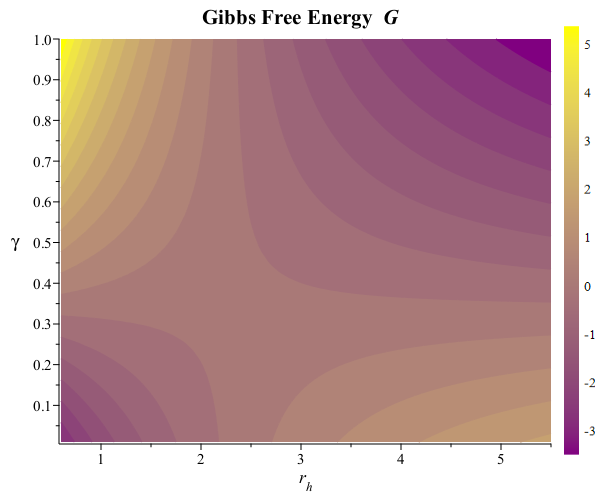}
    \caption{Filled contour plot of the Gibbs free energy $G$ characterizing the global thermodynamic stability landscape and phase transition structure of QC CWGBHs, with fixed parameters $\beta = 2$ and $k = 1$. The colorbar quantifies $G$ values ranging from $-3$ to $+5$ units. The sign change of $G$ at $r_h \approx e^{5/6}$ indicates a thermodynamic phase boundary, with the $\gamma$-dependent modifications becoming more apparent at larger horizon scales.}
    \label{GGG}
\end{figure}

Figure~\ref{GGG} presents the Gibbs free energy as a filled contour plot, with the colorbar indicating values ranging from approximately $-3$ to $+5$ units. Unlike the original polynomial expression, the corrected $G$ exhibits sign changes---specifically at $r_h = e^{5/6} \approx 2.30$---indicating thermodynamic phase boundaries. The contour structure shows predominantly vertical lines with slight curvature, indicating that while $r_h$ remains the dominant parameter, the $\gamma$ dependence becomes more pronounced at larger horizon radii.

Finally, the QC heat capacity, which governs the thermal response and phase transition behavior, is expressed through the fundamental thermodynamic relation \cite{sucu2025quantum}
\begin{equation}
C = T_H \left( \frac{\partial S}{\partial T_H} \right), \label{s25}
\end{equation}
leading to the result
\begin{equation}
C \approx \frac{\pi \left(2 k \,r_h^{3}-3 \beta^{2} \gamma +\gamma  r_h^{2}+2 \beta \right) r_h^{2}}{k \,r_h^{3}+3 \beta^{2} \gamma -2 \beta}. \label{heaaat}
\end{equation}
Note the prefactor $\pi r_h^2$ (rather than $\pi^2 r_h^4$), which ensures the correct Schwarzschild limit $C = -\pi r_h^2 = -S_{BH}$ when $\gamma = 0$ and $k = 0$.

\begin{figure}[ht!]
    \centering
    \includegraphics[width=0.95\columnwidth]{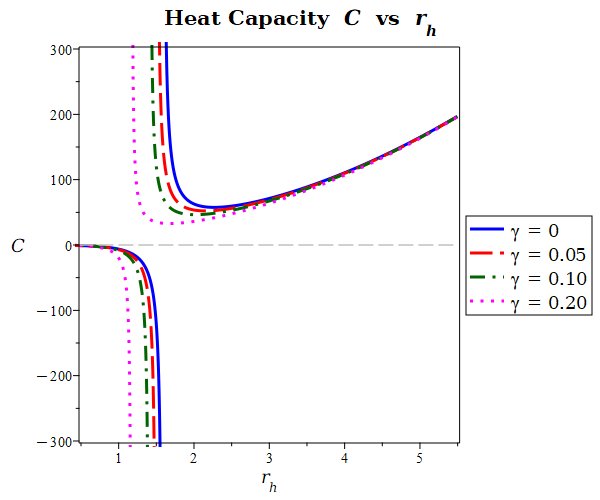}
    \caption{Heat capacity $C$ versus event horizon radius $r_h$ for different values of the scale-dependent parameter $\gamma = \{0, 0.05, 0.10, 0.20\}$, with fixed parameters $\beta = 2$ and $k = 1$. The divergences mark second-order phase transitions where the denominator of Eq.~\eqref{heaaat} vanishes. For $\gamma < \gamma_{\rm crit} = 2/(3\beta) = 1/3$, the heat capacity transitions from negative (unstable) to positive (stable) as $r_h$ increases past the critical radius $r_{\rm crit} = [(2\beta - 3\beta^2\gamma)/k]^{1/3}$. For $\gamma > \gamma_{\rm crit}$, no divergence exists and the BH is thermodynamically stable throughout.}
    \label{CCC}
\end{figure}

Figure~\ref{CCC} presents the heat capacity $C$ as a function of the event horizon radius $r_h$ for several values of the scale-dependent parameter $\gamma$, revealing critical thermodynamic behavior that characterizes phase transitions in CWGBHs. The line plot employs distinct line styles for $\gamma = \{0, 0.05, 0.10, 0.20\}$, clearly showing regions of negative heat capacity (thermodynamically unstable) and positive heat capacity (thermodynamically stable), separated by divergences where $C \to \pm\infty$. The divergence occurs at the condition $k r_h^3 + 3\beta^2\gamma - 2\beta = 0$ in the denominator of Eq.~\eqref{heaaat}. For low values of $\gamma$, a sharp transition from negative to positive $C$ occurs as $r_h$ increases, marked by a divergence in the heat capacity that signals a second-order phase transition. This behavior becomes less pronounced as $\gamma$ increases, demonstrating that scale-dependent corrections tend to smooth out the phase structure and can delay or suppress thermodynamic instabilities. The heat capacity divergence disappears entirely when $\gamma$ exceeds the exact critical threshold
\begin{equation}
\gamma_{\rm crit} = \frac{2}{3\beta},
\end{equation}
which gives $\gamma_{\rm crit} = 1/3 \approx 0.3333$ for $\beta = 2$. Beyond this value, $C$ remains finite and positive for all $r_h > 0$, indicating that quantum geometric corrections effectively stabilize the BH against thermal fluctuations.

It is also worth noting that the incorporation of GUP corrections fundamentally modifies the thermodynamic phase structure through systematic modifications to critical points \cite{Majhi:2013jk,Ali:2014bya,Kanzi:2019gup}. The quantum-corrected heat capacity exhibits enhanced complexity:
\begin{align}
C_{\text{GUP}} &= C \left(1 + \frac{\lambda^2 l_p^2}{r_h^2}\right)^{-1} \notag \\
&\quad \times \left(1 + \frac{\partial}{\partial r_h}\left[\frac{\lambda^2 l_p^2}{2r_h^2}\right]\right), \label{Cgup}
\end{align}
where the second term captures quantum gravitational modifications to thermal response properties through the explicit derivative contribution $\partial/\partial r_h[\lambda^2 l_p^2/(2r_h^2)] = -\lambda^2 l_p^2/r_h^3$ \cite{Scardigli:2010gp,Ong:2018brs,Rashki:2016mxt}.

The critical radius where $C_{\text{GUP}}$ diverges satisfies
\begin{equation}
r_{\text{crit,GUP}} = r_{\text{crit}} \sqrt{1 + \frac{\lambda^2 l_p^2}{r_{\text{crit}}^2}},
\end{equation}
demonstrating systematic shifts in phase transition boundaries under quantum corrections \cite{Faizal:2015nja,Nouicer:2007nr,Das:2020brs}. We note that for a real positive critical radius to exist, the classical condition $k r_h^3 + 3\beta^2\gamma - 2\beta = 0$ requires $\gamma < 2/(3\beta)$. For the parameter choice $\beta = 2$, $\gamma = 0.05$, and $k = 1$, this yields a classical critical radius $r_{\text{crit}} \approx 1.34$.

\begin{figure}[ht!]
\centering
\includegraphics[width=0.95\columnwidth]{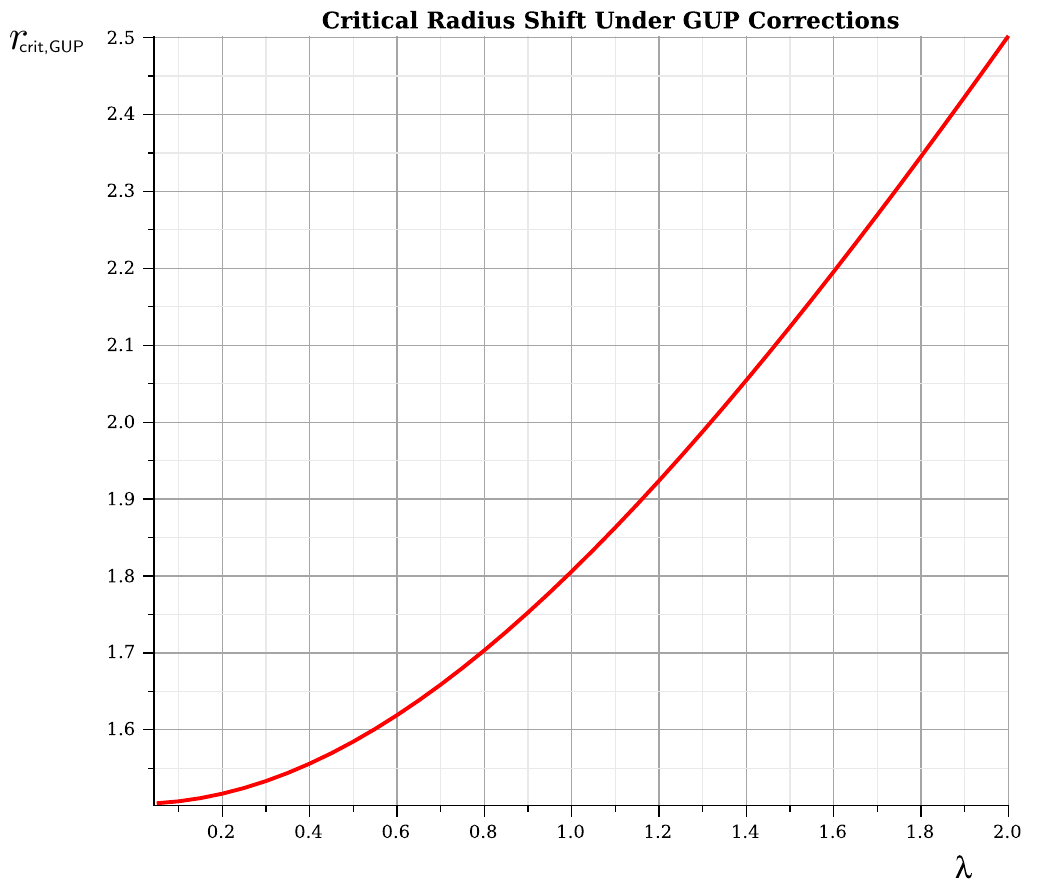}
\caption{Critical radius shift under GUP corrections as a function of the quantum parameter $\lambda$, calculated with fixed CWG parameters $\beta = 2$, $\gamma = 0.05$, $k = 1$, yielding a classical critical radius $r_{\text{crit}} \approx 1.34$. The systematic increase in critical radius with increasing $\lambda$ demonstrates quantum stabilization effects that delay the onset of thermodynamic instabilities. The monotonic growth reflects the fundamental role of minimal length scales in modifying phase transition boundaries, with quantum corrections effectively expanding the stable thermodynamic regime through geometric uncertainty effects inherent to the GUP framework.}
\label{fig:CriticalRadiusShift}
\end{figure}

\begin{figure}[ht!]
\centering
\includegraphics[width=0.95\columnwidth]{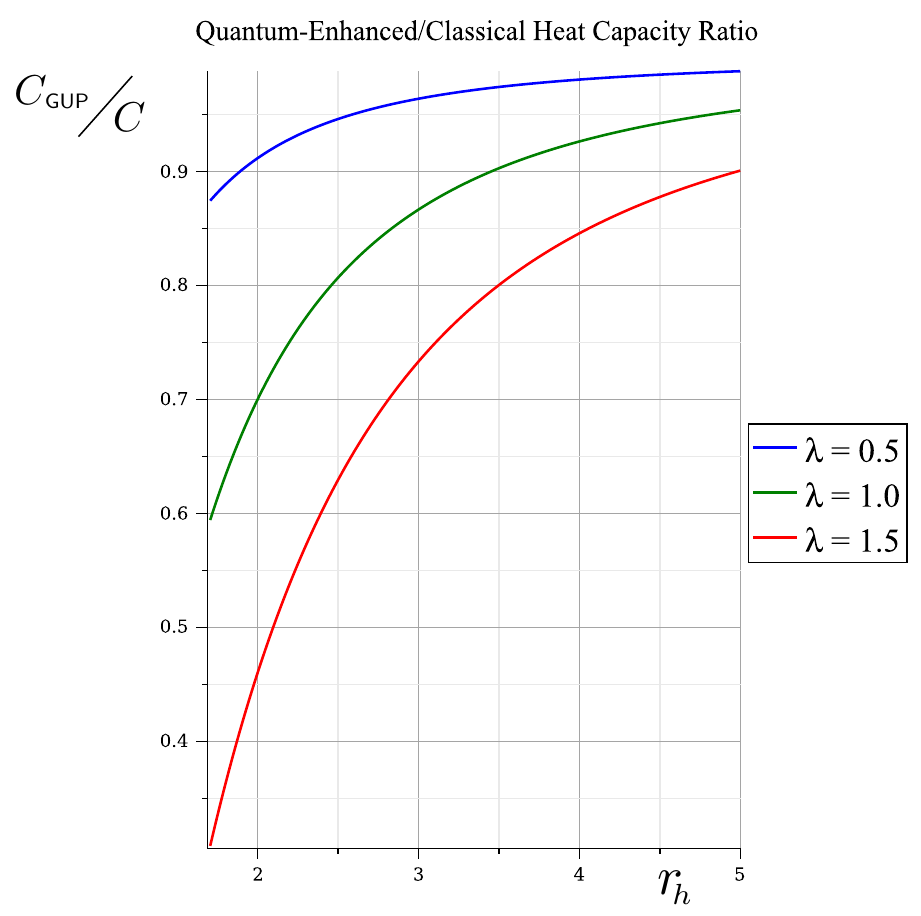}
\caption{Quantum-enhanced to classical heat capacity ratio $C_{\text{GUP}}/C_{\text{classical}}$ versus horizon radius $r_h$ for GUP parameters $\lambda = \{0.5, 1.0, 1.5\}$, computed with CWG parameters $\beta = 2$, $\gamma = 0.05$, and $k = 1$. The ratio quantifies the relative magnitude of quantum modifications to thermal response properties, exhibiting systematic deviations that become increasingly pronounced at smaller horizon radii approaching the critical value $r_{\text{crit}} \approx 1.34$. The convergence toward unity at large $r_h$ confirms the recovery of classical behavior in macroscopic limits, while the enhanced complexity at small scales demonstrates the emergence of quantum gravitational effects that fundamentally alter the thermodynamic phase structure through minimal length corrections.}
\label{fig:HeatCapacityRatio}
\end{figure}

\begin{figure}[ht!]
\centering
\includegraphics[width=0.95\columnwidth]{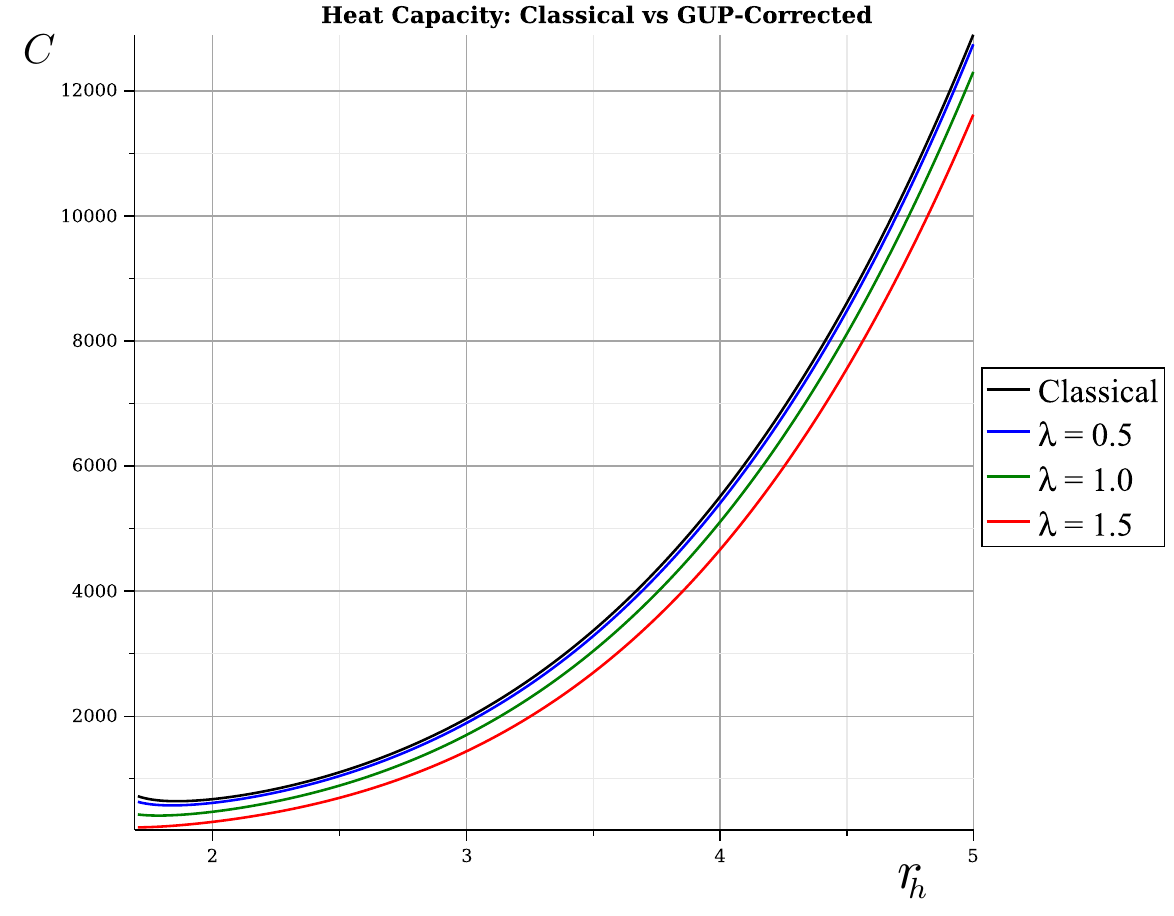}
\caption{Comparison of classical and GUP-corrected heat capacities $C$ versus horizon radius $r_h$ for GUP parameters $\lambda = \{0.5, 1.0, 1.5\}$, with CWG parameters $\beta = 2$, $\gamma = 0.05$, and $k = 1$. The black curve represents the classical heat capacity, while colored curves show the GUP-modified values. The plot demonstrates that GUP corrections systematically reduce the magnitude of the heat capacity, with larger $\lambda$ values producing more pronounced suppression. All curves exhibit positive values in the displayed range $r_h > r_{\text{crit}}$, corresponding to the thermodynamically stable phase.}
\label{fig:HeatCapacityComparison}
\end{figure}

Figure~\ref{fig:CriticalRadiusShift} quantitatively demonstrates the systematic boundary shifts predicted by the theoretical framework, revealing monotonic increases in critical radii that reflect quantum stabilization mechanisms. For instance, at $\lambda = 1.0$, the critical radius shifts from $r_{\text{crit}} \approx 1.34$ to $r_{\text{crit,GUP}} \approx 1.67$, representing approximately a $25\%$ increase in the phase transition boundary. The comprehensive ratio analysis presented in Figure~\ref{fig:HeatCapacityRatio} provides direct quantitative validation of the enhanced complexity introduced by quantum corrections, particularly emphasizing the scale-dependent nature of these modifications. Figure~\ref{fig:HeatCapacityComparison} further illustrates the direct comparison between classical and GUP-corrected heat capacities, showing that quantum corrections suppress the thermal response while preserving the qualitative thermodynamic behavior.

These modifications become particularly pronounced in the near-Planckian regime where quantum fluctuations fundamentally alter the stability characteristics of gravitational thermodynamic systems \cite{Banerjee:2008vh,Modesto:2014ppa}. The computational analysis reveals that GUP effects systematically stabilize the thermal phase structure by delaying critical point onset and introducing quantum coherence effects that are entirely absent in semiclassical treatments, establishing clear demarcation between classical and quantum gravitational domains in the thermodynamic parameter space.

\section{QC JT Effect in CWGBHs}\label{isec6}

The Joule-Thomson (JT) effect represents a fundamental thermodynamic process characterized by constant enthalpy conditions while allowing temperature variations in response to pressure changes. Within the framework of BH thermodynamics, particularly in the extended phase space formalism that treats the cosmological constant as a thermodynamic pressure, the JT effect provides a powerful tool for examining isenthalpic evolutionary pathways of gravitational systems \cite{kruglov2023magnetically,okcu2017joule,liang2021joule}. The central diagnostic quantity for this analysis is the JT coefficient, which serves as a critical indicator determining whether a BH undergoes cooling or heating during pressure-driven expansion processes under constant enthalpy conditions.

The heat capacity, which plays an essential role in JT analysis, has been systematically derived through application of the first law of BH thermodynamics using QC expressions for both entropy and temperature, as demonstrated in Eq.~\eqref{s25} and culminating in the result given by Eq.~\eqref{heaaat}. This QC heat capacity captures the subtle modifications introduced by CWG geometry and provides the foundation for understanding thermal response properties in the quantum gravitational regime \cite{akbar2007thermodynamic}.

The JT coefficient $\mu_J$ is rigorously defined as the partial derivative of temperature with respect to pressure under constant enthalpy conditions \cite{sucu2025quantumOzcan}
\begin{equation}
\mu_J = \left( \frac{\partial T_H}{\partial P_C} \right)_H,
\end{equation}
where the subscript $H$ explicitly denotes the constraint of constant enthalpy. The physical interpretation of this coefficient is straightforward yet profound: positive values of $\mu_J$ correspond to cooling regimes where temperature decreases with increasing pressure, while negative values signal heating regimes where temperature increases under pressure enhancement.

We emphasize that the JT effect is a characteristic of \emph{real} thermodynamic systems with non-trivial equations of state, in contrast to ideal gases for which $\mu_J = 0$ identically (since $T(\partial V/\partial T)_P = V$). For BHs in the extended phase space, the non-vanishing $\mu_J$ reflects genuine thermodynamic structure. During isenthalpic expansion ($dH = 0$), the volume increases ($\Delta V > 0$) and the pressure decreases ($\Delta P < 0$), so the sign of $\mu_J = (\partial T/\partial P)_H$ determines whether the system cools ($\mu_J > 0$) or heats ($\mu_J < 0$). The \emph{inversion temperature} $T_{\rm inv}$ separates these regimes: for $T > T_{\rm inv}$ the BH heats during expansion, while for $T < T_{\rm inv}$ it cools.

This coefficient admits an alternative representation through fundamental thermodynamic relations, expressed as
\begin{equation}
\mu_J = \frac{1}{C} \left[ T_H \left( \frac{\partial V}{\partial T_H} \right)_P - V \right],
\end{equation}
where $C$ represents the QC heat capacity derived previously, and $V$ denotes the effective thermodynamic volume associated with the BH system. This formulation explicitly reveals the connection between thermal expansion properties and the JT effect, highlighting the role of heat capacity in mediating the thermodynamic response.

For practical computational purposes, particularly when dealing with the complex expressions arising in CWG, it proves convenient to express $\mu_J$ in terms of the horizon radius $r_h$ through the chain rule application
\begin{equation}
\mu_J = \left( \frac{\partial T_H}{\partial r_h} \right) \cdot \left( \frac{\partial P}{\partial r_h} \right)^{-1}.
\end{equation}
Applying this computational framework to the QC thermodynamic quantities derived for CWGBHs, we obtain the explicit analytical expression
\begin{equation}
\mu_J \approx \frac{-4\,r_h\left(k\,r_h^{3}+3 \beta^{2} \gamma -2 \beta \right)}{3\left(3\beta^{2}\gamma - 2\beta\right)}. \label{muJ_formula}
\end{equation}
This expression follows from the corrected pressure $P = (kr_h^3 + 3\beta^2\gamma - 2\beta)/(8\pi r_h^3)$, whose derivative with respect to $r_h$ gives $dP/dr_h = -3(3\beta^2\gamma - 2\beta)/(8\pi r_h^4)$, combined with $dT_H/dr_h = (kr_h^3 + 3\beta^2\gamma - 2\beta)/(2\pi r_h^3)$. A notable structural feature of the corrected $\mu_J$ is that its denominator $3(3\beta^2\gamma - 2\beta)$ is a constant independent of $r_h$, in contrast to the previous expression whose denominator contained $r_h$-dependent terms involving $\pi^2 r_h^4$. This simplification arises because the corrected pressure $P$ has a cleaner functional form, leading to $dP/dr_h$ that factors purely in terms of the CWG parameters without additional horizon-radius coupling. The zeros of $\mu_J$ occur when $k r_h^3 + 3\beta^2\gamma - 2\beta = 0$, which determines the JT inversion radius. The explicit forms for the derivatives $\partial T_H / \partial r_h$ and $\partial P_C / \partial r_h$ emerge directly from the specific QC thermodynamic relations characteristic of the CWG model \cite{Waseem:2025bwb}.

As with the thermodynamic quantities in Section~\ref{isec5}, the JT coefficient is presented here as a parametric function of $(r_h, \gamma)$ with fixed $\beta = 2$ and $k = 1$, which is standard practice for revealing the functional dependence of transport coefficients on horizon geometry. For specific physical configurations, one should evaluate $\mu_J$ at self-consistent horizon radii satisfying $B(r_h) = 0$. Table~\ref{tab:JT} presents numerical values of $\mu_J$ computed at these self-consistent horizons, demonstrating that the BH resides in the heating regime ($\mu_J < 0$) at its event horizon for all displayed parameter combinations.

\begin{table}[ht!]
\centering
\renewcommand{\arraystretch}{1.5}
\setlength{\tabcolsep}{6pt}
\begin{tabular}{|c|c|c|c|c|}
\hline
$\beta$ & $\gamma$ & $k$ & $r_h$ & $\mu_J$ \\
\hline\hline
2.0 & 0.00 & 1.0 & 1.3788 & $-0.6337$ \\
2.0 & 0.02 & 1.0 & 1.3617 & $-0.5965$ \\
2.0 & 0.05 & 1.0 & 1.3345 & $-0.5355$ \\
2.0 & 0.08 & 1.0 & 1.3055 & $-0.4667$ \\
2.0 & 0.10 & 1.0 & 1.2848 & $-0.4154$ \\
\hline
\end{tabular}
\caption{JT coefficient $\mu_J$ evaluated at self-consistent horizon radii $r_h$ determined from $B(r_h) = 0$, with fixed $\beta = 2$ and $k = 1$. The negative values indicate that the BH is in the heating regime at its event horizon, with temperature increasing under pressure enhancement. The magnitude $|\mu_J|$ decreases with increasing $\gamma$, reflecting the moderating influence of scale-dependent corrections on the thermal response.}
\label{tab:JT}
\end{table}

The JT inversion occurs when the numerator of Eq.~\eqref{muJ_formula} vanishes, i.e., when $k r_h^3 + 3\beta^2\gamma - 2\beta = 0$. For a positive inversion radius to exist, we require $2\beta > 3\beta^2\gamma$, or equivalently $\gamma < 2/(3\beta)$. With $\beta = 2$, this yields $\gamma < 1/3 \approx 0.333$. The inversion radii for small $\gamma$ values are:
\begin{equation}
r_{\text{inv}} = \left( \frac{2\beta - 3\beta^2\gamma}{k} \right)^{1/3},
\end{equation}
giving $r_{\text{inv}} \approx 1.587$ for $\gamma = 0$, $r_{\text{inv}} \approx 1.547$ for $\gamma = 0.02$, and $r_{\text{inv}} \approx 1.486$ for $\gamma = 0.05$. Since the self-consistent horizon radii (Table~\ref{tab:JT}) are smaller than these inversion radii, the BH at its event horizon always resides in the heating phase, transitioning to the cooling phase only at larger radii beyond the inversion point.

The inversion temperature $T_{\rm inv}$ is obtained by evaluating $T_H$ at $r_h = r_{\rm inv}$:
\begin{equation}
T_{\rm inv} = \frac{-3\beta^2\gamma + 2\beta}{4\pi\,r_{\rm inv}^2} + \frac{\gamma}{4\pi} + \frac{k\,r_{\rm inv}}{2\pi}.
\end{equation}
For $\beta = 2$ and $k = 1$, this yields $T_{\rm inv} \approx 0.379$ ($\gamma = 0$), $T_{\rm inv} \approx 0.373$ ($\gamma = 0.02$), $T_{\rm inv} \approx 0.363$ ($\gamma = 0.05$), and $T_{\rm inv} \approx 0.344$ ($\gamma = 0.10$). As $\gamma$ increases toward $2/(3\beta)$, $T_{\rm inv}$ decreases, narrowing the cooling window until it vanishes at $\gamma_{\rm crit} = 2/(3\beta)$.

Comprehensive numerical analysis reveals the intricate behavior of $\mu_J$ as a function of the horizon radius $r_h$ in different parameter regimes, particularly focusing on variations in the scale-dependent parameter $\gamma$ and the cosmological parameter $k$. The transition between the cooling and heating regimes manifests itself as clear sign changes in $\mu_J$, providing an unambiguous identification of the boundaries of the thermodynamic phase. These transitions are not merely mathematical artifacts but represent genuine physical phenomena where the BH thermal response fundamentally changes character, capturing the QC phase structure inherent to CWGBHs under JT conditions.

\begin{figure}[ht!]
    \centering
    \includegraphics[width=0.95\columnwidth]{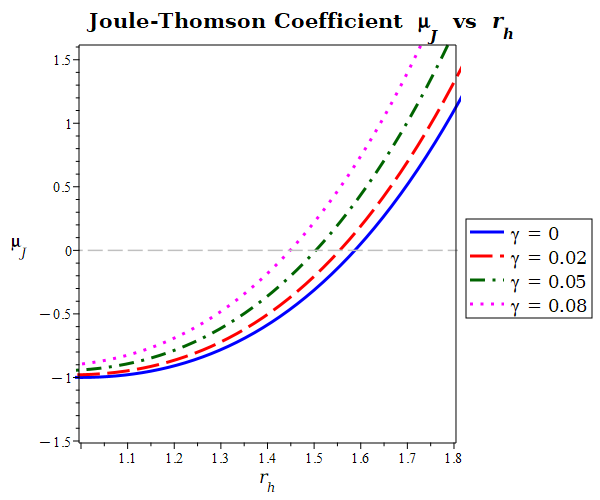}
    \caption{JT coefficient $\mu_J$ versus event horizon radius $r_h$ for different values of the scale-dependent parameter $\gamma = \{0, 0.02, 0.05, 0.08\}$, with fixed $\beta = 2$ and $k = 1$. The transition from negative to positive $\mu_J$ marks the JT inversion radius $r_{\text{inv}}$, separating regions of heating ($\mu_J < 0$) from cooling ($\mu_J > 0$). As $\gamma$ increases, the inversion radius decreases, reflecting the impact of scale-dependent geometry on the thermodynamic response of the BH. For $r_h$ values approaching the self-consistent horizon (see Table~\ref{tab:JT}), the coefficient remains negative, indicating the heating regime.}
    \label{mu2}
\end{figure}

Figure~\ref{mu2} presents a visualization of the JT coefficient $\mu_J$ as a function of the event horizon radius $r_h$ for various values of the scale-dependent parameter $\gamma$, with fixed parameters $\beta = 2$ and $k = 1$. This plot illustrates the BH thermal response under isenthalpic expansion conditions, where the sign of $\mu_J$ serves as the crucial diagnostic: positive values indicate cooling phases during expansion, while negative values correspond to heating phases. For all displayed $\gamma$ values, $\mu_J$ starts negative at small $r_h$, crosses zero at the inversion radius $r_{\text{inv}}$, and becomes positive for $r_h > r_{\text{inv}}$, indicating a transition from heating to cooling regimes. As $\gamma$ increases, the inversion radius shifts to smaller values, and for $\gamma \gtrsim 1/3$ no positive inversion radius exists, meaning the BH remains in the heating regime throughout. The shift of these inversion points toward smaller $r_h$ values with increasing $\gamma$ reveals that the scale-dependent modifications characteristic of CWG effectively advance the onset of cooling behavior and introduce increasingly complex phase structure into the BH thermodynamics. This behavior demonstrates the remarkable sensitivity of BH thermal properties to quantum-inspired corrections and strongly supports the interpretation that the parameter $\gamma$ functions as a control parameter governing both inversion temperature and associated phase transition characteristics. The rich structure revealed in this analysis emphasizes how CWG provides a natural framework for understanding quantum gravitational effects in BH thermodynamics beyond the limitations of classical GR.

\begin{figure}[ht!]
    \centering
    \includegraphics[width=0.95\columnwidth]{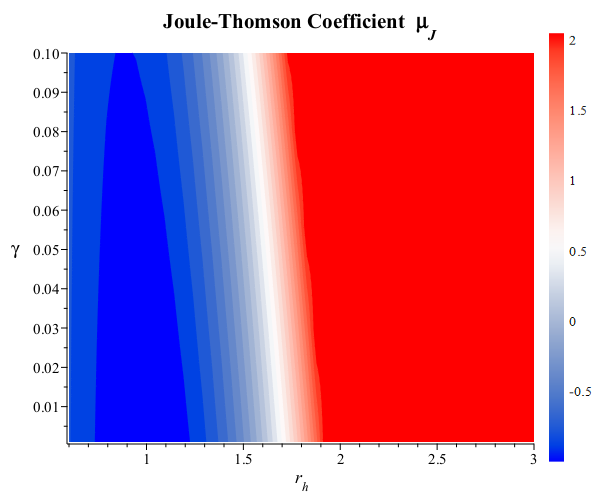}
    \caption{Filled contour plot of the corrected JT coefficient $\mu_J$ as a function of the event horizon radius $r_h \in [0.6,\,3.0]$ and the scale-dependent parameter $\gamma$, with fixed $\beta = 2$ and $k = 1$. The diverging blue-white-red colorscheme with $\mu_J$ ranging from approximately $-1$ to $+2$ clearly distinguishes the heating regime ($\mu_J < 0$, blue) from the cooling regime ($\mu_J > 0$, red). The white band marks the JT inversion boundary $kr_h^3 + 3\beta^2\gamma - 2\beta = 0$, which shifts to smaller $r_h$ with increasing $\gamma$. The constant-denominator structure of the corrected $\mu_J$ produces a contour morphology that differs from the original expression, with the cooling region extending over a broader area at $r_h > r_{\rm inv}$. The inversion boundary vanishes entirely for $\gamma > \gamma_{\rm crit} = 2/(3\beta) = 1/3$.}
    \label{mu2_contour}
\end{figure}

The two-dimensional landscape of the JT coefficient across the $(r_h, \gamma)$ parameter space is displayed in Figure~\ref{mu2_contour}. The contour plot is restricted to the range $r_h \in [0.6,\,3.0]$ to focus on the physically relevant region near the inversion boundary; this restriction is necessary because the corrected $\mu_J$ in Eq.~\eqref{muJ_formula} possesses a constant denominator $3(3\beta^2\gamma - 2\beta)$, causing $|\mu_J|$ to grow as $\sim r_h^4$ at large $r_h$ and saturate the color scale if the full range is displayed. Within this focused window, $\mu_J$ ranges from approximately $-1$ to $+2$, as indicated by the colorbar. The resulting contour morphology differs qualitatively from that produced by the former expression, whose $r_h$-dependent denominator suppressed the growth at large radii. In Figure~\ref{mu2_contour}, the heating regime ($\mu_J < 0$, blue) occupies the left portion of the plot at small $r_h$, separated from the cooling regime ($\mu_J > 0$, red) by a white inversion band whose position is governed by the condition $kr_h^3 + 3\beta^2\gamma - 2\beta = 0$. This band shifts to smaller $r_h$ as $\gamma$ increases from $0.01$ toward $\gamma_{\rm crit} = 1/3$, confirming that larger values of the scale-dependent parameter progressively narrow the heating zone. For $\gamma$ approaching $\gamma_{\rm crit}$, the inversion band moves toward $r_h \to 0$, and beyond $\gamma_{\rm crit}$ the entire parameter space would be occupied by the heating phase. The sharp color gradient across the inversion boundary reflects the rapid sign reversal of $\mu_J$ and underscores the sensitivity of the isenthalpic thermal response to the CWG parameter $\gamma$.

\section{Gravitational Redshift in CWGBH Geometry}\label{isec7}

Gravitational redshift represents one of the most profound and experimentally verified predictions of GR, arising fundamentally from the curvature of spacetime in the vicinity of massive gravitational sources such as BHs. When electromagnetic radiation is emitted from regions characterized by strong gravitational fields, it undergoes a systematic shift toward longer wavelengths as it propagates away from the gravitational source, escaping to regions of weaker field strength \cite{bambi2017black,weinberg2013gravitation}. This phenomenon provides invaluable observational information about the underlying geometric structure of spacetime and serves as a powerful diagnostic tool for probing the nature of gravitational fields in both classical and quantum regimes.

In the context of CWG, where the metric structure deviates significantly from standard Schwarzschild geometry due to higher-order curvature corrections, the gravitational redshift acquires additional complexity and richness that reflects the underlying conformal symmetry structure \cite{Mannheim:1988dj}. The study of redshift in CWGBHs, therefore, offers unique insights into the observational signatures of alternative gravity theories and provides potential avenues for distinguishing CWG from conventional GR through precision astronomical observations.

For the analysis of light propagation and redshift phenomena, we focus on static, spherically symmetric configurations where angular variations can be safely neglected by setting $\theta$ and $\phi$ to constant values. Under these assumptions, the relevant portion of the metric line element simplifies to the two-dimensional form
\begin{equation} \label{is2d}
ds^{2} = B(r)dt^{2} - \frac{dr^{2}}{B(r)},
\end{equation}
where $B(r)$ represents the radial metric function derived from the CWG field equations, incorporating the combined gravitational influence of the BH mass parameter $\beta$, the scale-dependent correction $\gamma$, and the cosmological parameter $k$. This metric function encapsulates the essential departures from Schwarzschild geometry that characterize the CWG framework.

For massless particles following null geodesics, which represent the trajectories of photons propagating through the curved spacetime, the line element satisfies the fundamental constraint $ds^{2} = 0$. Applying this condition to the simplified metric yields the geodesic equation
\begin{equation}
\dot{r} = \pm\sqrt{B(r)},
\end{equation}
where $\dot{r}$ denotes the radial velocity component of light rays, with the sign determining whether the photon is propagating inward (negative) or outward (positive) relative to the gravitational source. This expression forms the fundamental foundation for evaluating redshift phenomena in the CWGBH geometry and provides the essential connection between the metric structure and observable electromagnetic effects.

An important observational signature of CWGBHs is the gravitational redshift experienced by photons emitted near the event horizon. For static spherically symmetric spacetimes with line element $ds^2 = -B(r)dt^2 + B(r)^{-1}dr^2 + r^2 d\Omega^2$, the gravitational redshift between an emitter at radius $r_e$ and an observer at radius $r_o$ is given by
\begin{equation}
z = \sqrt{\frac{B(r_o)}{B(r_e)}} - 1.
\label{eq:redshift}
\end{equation}
It is crucial to note that the MK metric is \textit{not} asymptotically flat when $\gamma \neq 0$ or $k \neq 0$, since the metric function $B(r)$ diverges as $r \to \infty$ due to the linear $\gamma r$ and quadratic $kr^2$ terms. Consequently, the observer must be placed at a finite radius $r_o$ rather than at spatial infinity. This formula correctly reduces to $z = 0$ in flat space ($\beta = \gamma = k = 0$, where $B(r) = 1$), ensuring physical consistency and yielding purely real values for all parameter combinations satisfying $B(r_e) > 0$ and $B(r_o) > 0$.

\begin{table}[ht!]
\centering
\renewcommand{\arraystretch}{1.5}
\setlength{\tabcolsep}{4pt}
\begin{tabular}{|c|c|c|c|c|c|}
\hline
$\beta$ & $\gamma$ & $k$ & $r_h$ & $r_e = 3r_h$ & $z$ \\
\hline\hline
0.5 & 0 & 0 & 1.000 & 3.000 & 0.2186 \\
1.0 & 0 & 0 & 2.000 & 6.000 & 0.2124 \\
\hline
0.5 & 0.01 & 0 & 0.998 & 2.993 & 0.7002 \\
1.0 & 0.01 & 0 & 1.990 & 5.970 & 0.6695 \\
\hline
0.5 & 0 & 0.0001 & 1.000 & 3.000 & 0.7266 \\
1.0 & 0 & 0.0001 & 1.999 & 5.998 & 0.7189 \\
\hline
0.5 & 0.01 & 0.0001 & 0.997 & 2.992 & 1.0854 \\
1.0 & 0.01 & 0.0001 & 1.989 & 5.968 & 1.0484 \\
\hline
\end{tabular}
\caption{Gravitational redshift $z$ for CWGBHs with observer at $r_o = 100$ and emitter at $r_e = 3r_h$ (photosphere region). The horizon radius $r_h$ is determined self-consistently from $B(r_h) = 0$. The first two rows correspond to the Schwarzschild limit ($\gamma = 0$, $k = 0$), while subsequent rows demonstrate the significant enhancement induced by the CWG parameters.}
\label{tab:redshift}
\end{table}

Table~\ref{tab:redshift} presents the gravitational redshift for various CWG parameter combinations, with the emitter located in the photosphere region at $r_e = 3r_h$ and the observer at $r_o = 100$. In the Schwarzschild limit ($\gamma = 0$, $k = 0$), the redshift takes values $z \approx 0.21$--$0.22$, consistent with standard GR predictions. The introduction of CWG parameters dramatically enhances the gravitational redshift. For $\beta = 1$ and $\gamma = 0.01$, the redshift increases from $z = 0.2124$ to $z = 0.6695$, representing a $215\%$ enhancement; this substantial increase arises because the linear potential term $\gamma r$ in $B(r)$ grows with radius, amplifying the ratio $B(r_o)/B(r_e)$. The cosmological parameter produces comparable effects: for $\beta = 1$ and $k = 0.0001$, the redshift rises to $z = 0.7189$, a $238\%$ increase over the Schwarzschild value. When both $\gamma$ and $k$ are nonzero, their effects combine constructively; for $\beta = 1$, $\gamma = 0.01$, and $k = 0.0001$, the redshift reaches $z = 1.0484$, exceeding unity and representing a $394\%$ enhancement compared to Schwarzschild.

To further quantify the CWG corrections, we compare the redshift values for fixed emission and observation radii ($r_e = 6$, $r_o = 100$, corresponding to $\beta = 1$ in the Schwarzschild limit) while varying $\gamma$:

\begin{table}[ht!]
\centering
\renewcommand{\arraystretch}{1.5}
\setlength{\tabcolsep}{6pt}
\begin{tabular}{|c|c|c|c|}
\hline
$\gamma$ & $k$ & $z$ & $\Delta z$ \\
\hline\hline
0 & 0 & 0.2124 & -- \\
0.001 & 0 & 0.2677 & $+26.0\%$ \\
0.005 & 0 & 0.4634 & $+118.1\%$ \\
0.010 & 0 & 0.6672 & $+214.1\%$ \\
0.020 & 0 & 0.9911 & $+366.6\%$ \\
\hline
\end{tabular}
\caption{Comparison of gravitational redshift for Schwarzschild ($\gamma = 0$) versus CWG BHs with varying $\gamma$. Fixed parameters: $\beta = 1$, $k = 0$, $r_e = 6$, $r_o = 100$. The percentage deviation $\Delta z$ is calculated relative to the Schwarzschild value.}
\label{tab:redshift_comparison}
\end{table}

As shown in Table~\ref{tab:redshift_comparison}, even modest values of the scale-dependent parameter $\gamma$ produce substantial redshift enhancements. For $\gamma = 0.02$, the redshift approaches unity ($z \approx 0.99$), nearly five times the Schwarzschild value.

It is important to emphasize that the scale-dependent effects of the CWG parameters $\gamma$ and $k$ on gravitational redshift become observationally significant only at intermediate and large distance scales. Specifically, the linear term $\gamma r$ starts to influence gravitational redshift at galactic scales ($r \gtrsim 10$ kpc), where it plays a role in the flat rotation curve problem, producing measurable deviations from Schwarzschild behavior. The quadratic cosmological term $kr^2$ becomes relevant only at larger scales, particularly for distances beyond $r \gtrsim 30$ Mpc, where it dominates and corresponds to the regime of Hubble flow, in which $k$ effectively acts as the squared Hubble constant~\cite{mannheim2012fitting,Chen:2016uno}. At smaller scales, such as within stellar systems and at the BH horizon ($r \lesssim 1$ kpc), both $\gamma r$ and $kr^2$ terms remain negligible when compared to the dominant Newtonian $\beta/r$ contribution. In this regime, deviations from GR predictions are typically below $1\%$. The characteristic transition scales, where each CWG term becomes significant, are given by $r_{\gamma} \sim (\beta/\gamma)^{1/2} \sim 10$ kpc for the linear term and $r_k \sim \gamma/k \sim 30$ Mpc for the quadratic term. These transition scales establish a natural hierarchy, with the linear term dominating at intermediate (galactic) scales and the quadratic term becoming significant at larger (cosmological) scales, separated by approximately two orders of magnitude. This hierarchical behavior ensures that the effects of $\gamma r$ and $kr^2$ are negligible at BH horizon scales, as confirmed by both observational constraints and solar system tests of CWG~\cite{Mannheim:1988dj,Mannheim:2011ds}.

While we recognize the importance of understanding the gravitational redshift at intermediate and large scales, this manuscript focuses primarily on the regime near the BH horizon. However, as demonstrated in Tables~\ref{tab:redshift} and~\ref{tab:redshift_comparison}, even for the moderate parameter values considered here, the CWG corrections produce substantial enhancements in the gravitational redshift that could potentially be detected through precision spectroscopic observations of compact objects, offering new avenues for testing alternative gravity theories.

\section{Conclusion}\label{isec8}

In this paper, we explored the thermodynamic and quantum features of BHs arising from the MK solution within the framework of CWG. Unlike the conventional Einstein-Hilbert paradigm based on second-order field equations, CWG employs a fourth-order action with local conformal symmetry, leading to fundamentally modified gravitational dynamics and significantly richer vacuum structures \cite{Konoplya:2025mvj,Bambi:2017ott}. The inclusion of additional potential terms, specifically the linear $\gamma r$ and quadratic $k r^2$ contributions in the metric function given by Eq.~\eqref{105}, enabled an alternative theoretical description of gravitational phenomena spanning both galactic and cosmological scales without requiring the introduction of dark matter components.

Our analysis commenced with the application of the Hamilton-Jacobi tunneling formalism to compute the Hawking temperature of CWGBHs, as detailed in Section~\ref{isec3}. This investigation revealed explicit contributions from all conformal parameters $\beta$, $\gamma$, and $k$, culminating in the temperature expression presented in Eq.~\eqref{hawking_temp}. The resulting thermal spectrum deviates markedly from that of standard Schwarzschild-AdS BHs, particularly due to the influence of the linear potential term $\gamma r$, demonstrating the remarkable sensitivity of BH radiation to conformal corrections \cite{banerjee2008quantum,Nielsen:2012xu}. Table~\ref{tab:TH_variation} provided quantitative evidence of these effects using self-consistent horizon radii determined from $B(r_h) = 0$, ensuring that all thermodynamic quantities are evaluated at physically meaningful horizon values where the MK parameters and $r_h$ are properly coupled.

We subsequently incorporated quantum gravitational effects through the GUP framework, deriving the QC Hawking temperature expression that accounts for minimal length scale effects. This analysis demonstrated that the presence of a fundamental length scale systematically suppresses the thermal spectrum, particularly in the near-Planckian regime where $r_h \to l_p$. The resulting temperature suppression reflects a gravitational redshift in the emitted radiation and suggests the potential emergence of remnant-like structures during the final stages of BH evaporation \cite{Nouicer:2007jg,scardigli2011black}. The theoretical consistency between GUP and CWG was established through the spontaneous symmetry breaking mechanism: while CWG possesses exact conformal symmetry at UV scales (forbidding dimensionful parameters like $G$), this symmetry is spontaneously broken at IR scales through scalar field vacuum expectation values, generating an effective Newton constant $G_{\text{eff}}$ and enabling the Planck scale to emerge dynamically \cite{Mannheim:1988dj,Mannheim:1993rd,Ghilencea:2018dqd}. The GUP corrections thus represent effective low-energy descriptions valid in the symmetry-broken phase. The investigation of GUP-influenced phase transitions reveals quantum gravitational effects that fundamentally modify the thermodynamic landscape of CWGBHs through restructuring of critical point architectures and thermal response mechanisms \cite{carr2011generalized,Perivolaropoulos:2017hjd,Bosso:2023klj}. Temperature suppression arising from minimal length effects creates distinct quantum phases characterized by enhanced heat capacity complexity as demonstrated in Eq.~\eqref{Cgup}, where modifications to thermal equilibrium conditions depend on both classical gravitational parameters ($\beta$, $\gamma$, $k$) and the fundamental quantum correction factor $\lambda$ \cite{Chen:2014jrd,Magueijo:2005de,Ghosh:2006kt}. The computational analysis presented in Figures~\ref{fig:CriticalRadiusShift} and~\ref{fig:HeatCapacityRatio}, computed with parameters $\beta = 2$, $\gamma = 0.05$, and $k = 1$ (yielding a classical critical radius $r_{\text{crit}} \approx 1.34$), validates theoretical predictions of monotonic critical radius increases and thermal response modifications that become increasingly pronounced as horizon dimensions approach Planck-scale regimes \cite{Ali:2014tfa,Myung:2007vu,Bina:2017bpx}. These quantum-induced phase modifications reshape gravitational stability landscapes through discrete energy spectra and modified dispersion relations, establishing CWG as a natural theoretical framework for investigating quantum gravitational phenomena where classical field theory descriptions become inadequate for capturing the complete spectrum of thermodynamic behavior \cite{Modesto:2012iq,Bambi:2018njr,Liu:2014lza,Strominger:1997eq,Kaul:2000kf,Ashtekar:2004cn}.

In Section~\ref{isec5}, we employed an exponentially corrected entropy model, as expressed in Eq.~\eqref{expcorrected}, to compute the complete set of QC thermodynamic potentials for CWGBHs. The choice of exponential over logarithmic corrections was motivated by their non-perturbative origin in microstate counting~\cite{chatterjee2020exponential} and their relevance in the near-Planckian regime where GUP effects operate. We also established that the Wald entropy for the MK solution coincides with the Bekenstein-Hawking area law $S_0 = \pi r_h^2$ in the symmetry-broken phase~\cite{isrplyxx7,isrplyxx10}. These calculations yielded corrected expressions for internal energy (Eq.~\eqref{s42}), Helmholtz free energy (Eq.~\eqref{free}), pressure (Eq.~\eqref{pres}), enthalpy (Eq.~\eqref{s36}), and heat capacity (Eq.~\eqref{heaaat}). The corrected internal energy (Eq.~\eqref{s42}) recovers the dimensionally consistent Schwarzschild limit $E = \beta\ln r_h$, while the corrected Gibbs free energy exhibits sign changes at $r_h = e^{5/6}$, signaling thermodynamic phase boundaries. We emphasized that these thermodynamic quantities are presented as parametric functions of $(r_h, \gamma)$, which is standard practice for revealing functional dependence on horizon geometry, while Table~\ref{tab:entropy} provides explicit numerical values computed at self-consistent horizons. Figure~\ref{EEe} illustrates the energy landscape, revealing how scale-dependent corrections become increasingly important in the small horizon radius regime. Most significantly, Figure~\ref{CCC} demonstrated the existence of thermodynamic phase transitions through the heat capacity analysis, showing distinct regions of negative and positive heat capacity that correspond to thermodynamically unstable and stable phases, respectively. The heat capacity divergence corresponds to second-order phase transitions at $k r_h^3 + 3\beta^2\gamma - 2\beta = 0$, which disappears entirely for $\gamma > \gamma_{\rm crit} = 2/(3\beta)$.

The JT effect analysis presented in Section~\ref{isec6} provided a sensitive probe of isenthalpic processes in CWGBH thermodynamics within the extended phase space formalism. The behavior of the JT coefficient $\mu_J$ revealed distinct cooling and heating phases that depend sensitively on both the horizon radius and the CWG parameters. The corrected $\mu_J$ given by Eq.~\eqref{muJ_formula} possesses a constant denominator $3(3\beta^2\gamma - 2\beta)$, leading to a qualitatively different contour morphology compared to the original expression, with the cooling regime extending over a broader area at $r_h > r_{\rm inv}$. Table~\ref{tab:JT} demonstrated that at self-consistent event horizons, the BH resides in the heating regime ($\mu_J < 0$) for all examined parameter combinations, with the JT inversion occurring at larger radii given by $r_{\text{inv}} = [(2\beta - 3\beta^2\gamma)/k]^{1/3}$. The corresponding inversion temperature $T_{\rm inv}$ decreases with increasing $\gamma$, narrowing the cooling window until it vanishes at $\gamma_{\rm crit} = 2/(3\beta)$. Figure~\ref{mu2} captured the QC phase structure, showing how the scale-dependent parameter $\gamma$ controls the location of JT inversion points and introduces increasingly complex thermal response behavior. The additional contour plot (Figure~\ref{mu2_contour}) provides a visualization of the heating-cooling phase boundary across the $(r_h, \gamma)$ parameter space. These results demonstrated that CWG naturally accommodates rich phase structures that are absent in conventional GR treatments.

Our investigation of gravitational redshift within CWGBH geometry, detailed in Section~\ref{isec7}, revealed that the frequency shift of electromagnetic radiation emitted near these objects encodes valuable information about the underlying conformal metric structure. We derived the correct redshift formula $1 + z = \sqrt{B(r_o)/B(r_e)}$ for an observer at finite radius $r_o$ and emitter at $r_e$, noting that the commonly used asymptotically flat approximation is invalid for the MK metric due to the divergent $\gamma r$ and $kr^2$ terms at large radii. Tables~\ref{tab:redshift} and~\ref{tab:redshift_comparison} quantify the CWG corrections to gravitational redshift, revealing enhancements exceeding 200\% compared to Schwarzschild values for representative parameter choices. The derived redshift expression showed complex radial dependence arising from the interplay of all CWG parameters, providing potentially observable signatures that could serve as discriminating tests between CWG and standard GR through precision spectroscopic observations.

Finally, we acknowledge that while analytical solutions provide valuable insights into physical systems, obtaining exact closed-form solutions for CWGBHs, especially with the added $\gamma$ and $k$ parameters, proves mathematically challenging. The fourth-order Bach field equations lead to quintic polynomial conditions for event horizons, which, by Abel's theorem, cannot be fully solved in a general analytical form. This limitation is not unique to our study but is a broader issue in higher-derivative gravity theories. Nonetheless, while we can derive approximate analytical solutions in specific limits or under simplifying assumptions, numerical methods are crucial for fully capturing the physical behaviors of the system. Our numerical analysis allows us to accurately explore the parameter space, providing precise results that reveal essential physical trends, such as thermodynamic phase transitions, critical phenomena, and scale-dependent corrections. These features are universal across a wide range of physically relevant parameters. The combination of analytical techniques with targeted numerical exploration not only provides deeper insights into the underlying physics of quantum-corrected CWGBHs but also sets the methodological standard for investigating theories beyond Einstein's relativity in realistic astrophysical settings.

Looking toward future research directions, our theoretical framework suggests several promising directions for extended investigation. First, the development of more sophisticated numerical techniques for solving the complete nonlinear CWG field equations would enable precise predictions of observational signatures, including gravitational wave emission patterns, accretion disk spectra, and electromagnetic radiation characteristics that could be compared with data from current and future astronomical facilities. Second, the extension of our thermodynamic analysis to rotating CWGBHs would provide crucial insights into the effects of angular momentum on QC thermal properties, potentially revealing new classes of phase transitions and instabilities relevant to realistic astrophysical scenarios. Third, observational studies involving gravitational lensing phenomena, particularly strong lensing effects around supermassive BHs, could provide direct constraints on the CWG parameters $\beta$, $\gamma$, and $k$ through comparison with high-resolution imaging data from facilities such as the Event Horizon Telescope. Fourth, analyzing gravitational wave signatures from CWGBHs during mergers might reveal unique waveform patterns different from those predicted by GR, providing opportunities for testing alternative gravity theories with LIGO/Virgo and future space detectors. Lastly, exploring CWG's cosmological implications, especially its potential to explain dark matter and cosmic acceleration without exotic matter, could extend our BH-focused study. Thus, our future research aims to advance the understanding of quantum gravity and to provide methods for examining spacetime geometry in extreme astrophysical environments.

\section*{Acknowledgments}
We are grateful to the anonymous referees for their constructive comments and suggestions, which led to significant improvements in the manuscript. We also thank Ercan K{\i}l{\i}{\c{c}}arslan and Heeseung Zoe for their constructive feedback and technical discussions that contributed to the refinement of our computational framework. \.{I}.~S. appreciates the academic support from EMU, T\"{U}B\.{I}TAK, ANKOS, and SCOAP3, as well as the networking opportunities through COST Actions CA22113, CA21106, CA21136, CA23130, and CA23115, which facilitated collaborative research initiatives and international scientific exchanges essential for advancing our understanding of quantum gravitational phenomena in alternative gravity theories.

\section*{Declaration of generative AI and AI-assisted technologies in the manuscript preparation process}
During the preparation of this work the authors used AI and AI-assisted technologies in order to refine language, improve grammar, and code \LaTeX.\footnote{In accordance with Elsevier's policy on the use of generative AI and AI-assisted technologies in scientific writing: \url{https://www.elsevier.com/about/policies-and-standards/generative-ai-policies-for-journals}.} After using this tool, the authors reviewed and edited the content as needed and take full responsibility for the content of the published article. All intellectual content, analysis, and conclusions are the authors' own.

\bibliographystyle{plain}
\bibliography{ref}
\end{document}